\documentclass[aps,prb,twocolumn,showpacs,preprintnumbers,superscriptaddress,amsmath,amssymb,longbibliography,floatfix,10pt]{revtex4-2}
\usepackage{placeins}
\usepackage{graphicx}
\usepackage{dcolumn}
\usepackage{bm}
\usepackage{amsfonts}
\usepackage{amsmath}
\usepackage{amssymb}
\usepackage{color}
\usepackage{enumerate}
\usepackage{colortbl}
\usepackage[utf8]{inputenc}

\DeclareUnicodeCharacter{2212}{-}
\DeclareUnicodeCharacter{0301}{´}

\usepackage[dvipsnames]{xcolor}
\usepackage{ulem}

\begin{document}

\title{Correlation-enhanced magnetism at a superconducting cuprate–manganite interface}

\author{Vitor A. M. Lima}
\affiliation{Departamento de Física, Universidade Federal de Minas Gerais, C.P. 702, 30123-970 Belo Horizonte, Minas Gerais, Brazil}
\affiliation{Université Paris-Saclay, CNRS, Laboratoire de Physique des Solides, 91405, Orsay, France}
\author{Marcello Civelli}
\affiliation{Université Paris-Saclay, CNRS, Laboratoire de Physique des Solides, 91405, Orsay, France}
\author{Walber H. Brito}
\affiliation{Instituto de F\'{i}sica, Universidade de S\~ao Paulo, 05508-090 São Paulo, Brazil}

\date{\today}

\begin{abstract}
Motivated by experimental studies on cuprate-manganite heterostructures, we study a correlated bilayer Hubbard model subject to a local interfacial exchange field using cluster dynamical mean-field theory in the underdoped cuprate regime. We find that superconductivity coexists with a manganite-induced Cu spin polarization. Remarkably, throughout this coexistence regime, the superconducting solution carries a larger magnetic moment than the corresponding constrained normal state. Analysis of the magnetic response shows that this reversal is driven by correlations, which enhance magnetism already in the normal state and even more strongly in the superconducting solution. At stronger exchange fields, superconductivity collapses abruptly, displaying hysteresis consistent with a first-order transition, and a strongly spin-polarized pseudogap emerges. These results show that, under strong electronic correlations, superconductivity can enhance rather than suppress interfacial magnetism.
 
\end{abstract}

\maketitle


\begin{figure*}[t!]
    \centering
    \includegraphics[width=0.9\linewidth]{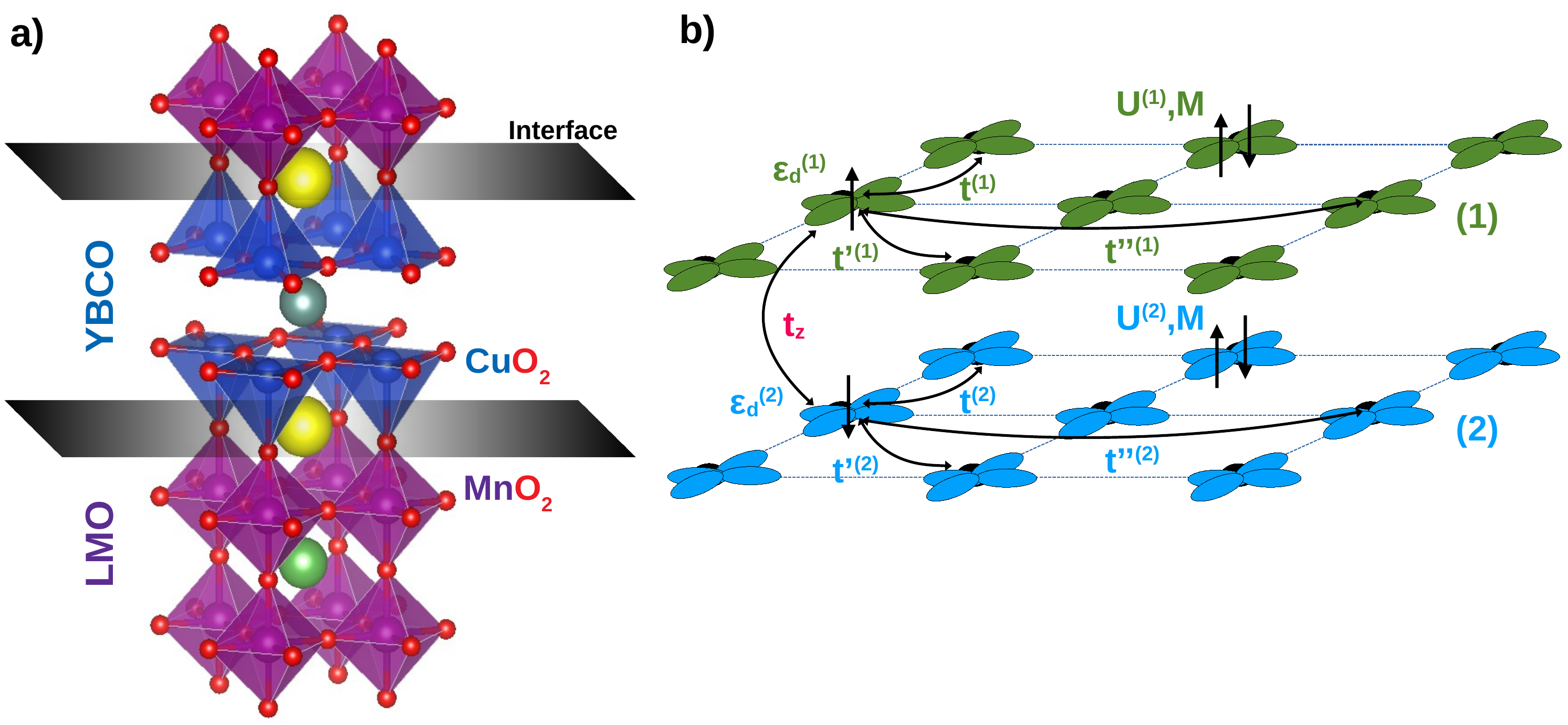}
    \caption{\textbf{a)} Relaxed structure of YBa$_2$Cu$_3$O$_7$ (1 u.c.)/LaMnO$_3$ (1 u.c.) reference supercell, with MnO$_2$--BaO interface termination. Blue, violet, and red spheres represent the Cu, Mn, and O atoms, respectively. Ba atoms are represented in yellow, while La in green. \textbf{b)} Effective bilayer Hubbard model for the two CuO$_2$ planes. We retain hopping amplitudes up to third-nearest neighbors ($t$, $t'$, and $t''$). The two planes are coupled by the interlayer hopping $t_z$, while local correlations and on-site energies are described by $U$ and $\varepsilon_d$, respectively. In the schematic, $M$ labels the spin-dependent on-site term associated with magnetic proximity to the manganite; it shifts the two spin projections in opposite directions by $\pm M$, producing an exchange splitting $2M$.}
    \label{fig:struct_param}
\end{figure*}

\section{\label{sec:int}Introduction} 

Cuprate--manganite interfaces provide a natural platform to investigate the interplay between proximity-induced ferromagnetism and superconductivity within the same CuO$_2$ electronic environment. When both orders act on the same orbital degrees of freedom, they may give rise to unconventional interfacial phases. Bulk cuprates, such as YBa$_2$Cu$_3$O$_7$ (YBCO), are spin-singlet superconductors with $d_{x^2-y^2}$ pairing symmetry~\cite{ybco_dx2y2}. By contrast, a ferromagnetic interface naturally favors spin-triplet correlations, allowing singlet and triplet pairing channels to coexist or become intertwined~\cite{bergeret_volkov_efetov2005,buzdin2005}. Remarkably, polarized neutron reflectometry on a cuprate--manganite superlattice revealed a large superconductivity-induced modulation of the ferromagnetic magnetization profile below $T_c$~\cite{Hoppler_2009}, demonstrating that the feedback of superconductivity on interfacial magnetism can be highly nontrivial. Despite previous experimental and theoretical efforts, the nature of the superconducting state at ferromagnetic cuprate/manganite interfaces, and its possible coexistence with interfacial magnetism, remains an important open question.

Here, we address the question of whether $d$-wave superconductivity can coexist with a manganite-induced Cu spin polarization within the same CuO$_2$ layer. The coexistence regime also contains an induced odd-frequency $S_z=0$ spin-triplet component~\cite{berezinskii1974,linder_balatsky2019,bergeret_volkov_efetov2005}. Our central result, however, concerns the magnetic response: throughout this regime, the superconducting solution carries a larger magnetic moment than the corresponding constrained normal state at the same hole density and exchange field. Analysis of the response shows that this counterintuitive reversal is driven by strong correlations, which enhance magnetism already in the normal state and even more strongly in the superconducting solution. At stronger exchange fields, superconductivity collapses abruptly, displaying hysteresis consistent with a first-order transition, and a strongly spin-polarized pseudogap emerges. Viewed from the high-field side, the spectral evolution suggests that the superconducting state, with its anomalously large magnetic moment, develops out of this correlated, spin-polarized pseudogap background. The correlation-enhanced magnetic response and the emergence of the spin-polarized pseudogap constitute the two principal results of this work.

From a materials perspective, advances in atomic-scale growth and characterization tools have enabled the exploration of transition-metal-oxide heterostructures, where the interplay of charge, spin, orbital, and lattice degrees of freedom can generate novel interfacial states. Representative examples include the two-dimensional electron gas at the SrTiO$_3$/LaAlO$_3$ interface~\cite{gas_2d} and emergent ferromagnetism in LaNiO$_3$/La$_{2/3}$Ca$_{1/3}$MnO$_3$ superlattices~\cite{Soltan2023}. In cuprate-based heterostructures, strong electronic correlations introduce a rich competition among different electronic orders, while charge transfer, confinement, and interfacial coupling provide additional tuning mechanisms. Early studies of interfaces between YBCO and La$_{2/3}$Ca$_{1/3}$MnO$_3$ (LCMO) or La$_{2/3}$Sr$_{1/3}$MnO$_3$ (LSMO) showed that the presence of Mn atoms close to the CuO$_2$ planes gives rise to small uncompensated magnetic moments at the interfacial copper atoms~\cite{dead_layer}. The magnetic anisotropy and texture of these heterostructures depend sensitively on layer thickness, interface termination, and sample geometry. In-plane Cu moments with ferromagnetic correlations extending along the crystallographic $c$ axis have been reported in YBCO/LCMO bilayers~\cite{selective_interlayer_coup}. Magnetization and ferromagnetic-resonance studies further reveal a thickness-dependent onset of ferromagnetism, a heterogeneous interfacial anisotropy, and perpendicular stray-field components generated by the magnetic texture~\cite{Dybko_2013,Uspenskaya_2021,Chaudhuri_2023}. Charge transfer between the materials was also reported~\cite{xas_chg_transf}. Unconventional magnetic and transport responses have further been interpreted as signatures of triplet proximity correlations\cite{tri_cup_man_1,CHOU2024158739,Sanchez-Manzano2022,Kumawat2023}. In addition, transport and spectroscopy measurements revealed a thickness-driven superconductor-to-insulator transition~\cite{sic_ybco_lsmo}, as well as putative two-dimensional altermagnetism~\cite{tri_cup_man_3}.

A previous theoretical study of the electronic and magnetic properties of the normal phase~\cite{meu_int_dft}, performed by some of the authors, revealed a relationship between the YBCO and LSMO thickness ratio and both the charge transfer and the magnetic moments induced at the interfacial cuprate layers in films with fewer than 3 unit cells (u.c.) of YBCO and 6 unit cells of LSMO. The density functional theory (DFT)+U results indicate that a decrease in the thickness ratio (\textit{i.e.}, an increase in the LSMO thickness) induces a larger local magnetic moment in the first cuprate layers, together with an enhanced charge transfer across the interface. These mechanisms are largely decoupled: magnetization is a short-range effect associated with interfacial bonding, whereas charge transfer extends over longer length scales. In all cases studied, the most stable magnetic configuration in the CuO$_2$ plane corresponds to a ferromagnetic ordering.

Building on this previous normal-state analysis, here we explicitly incorporate strong electronic correlations in the CuO$_2$ layers using cluster dynamical mean-field theory~\cite{kotliar_cdmft_prl,maier_cluster_rmp,qcm} (CDMFT). We focus on an interfacial YBCO bilayer corresponding to the two CuO$_2$ planes closest to the manganite layer. Within this framework, we show that correlations form robust local moments already in the underlying normal state. Counterintuitively, these pre-existing moments are more strongly polarized in the superconducting solution, producing a magnetic response beyond the quadratic BCS expectation.

The paper is organized as follows: 
In Sec.~\ref{sec:mod}, we introduce the bilayer Hubbard model and the CDMFT approach used to describe cuprate--manganite interfaces. Our results on the interplay between ferromagnetism and superconductivity, as well as on the polarized pseudogap phase, are presented in Sec.~\ref{sec:res}. The conclusions and outlook are presented in Sec.~\ref{sec:conc}.

\section{\label{sec:mod}Model and Methods}

\begin{figure}
    \centering
    \includegraphics[width=0.85\linewidth]{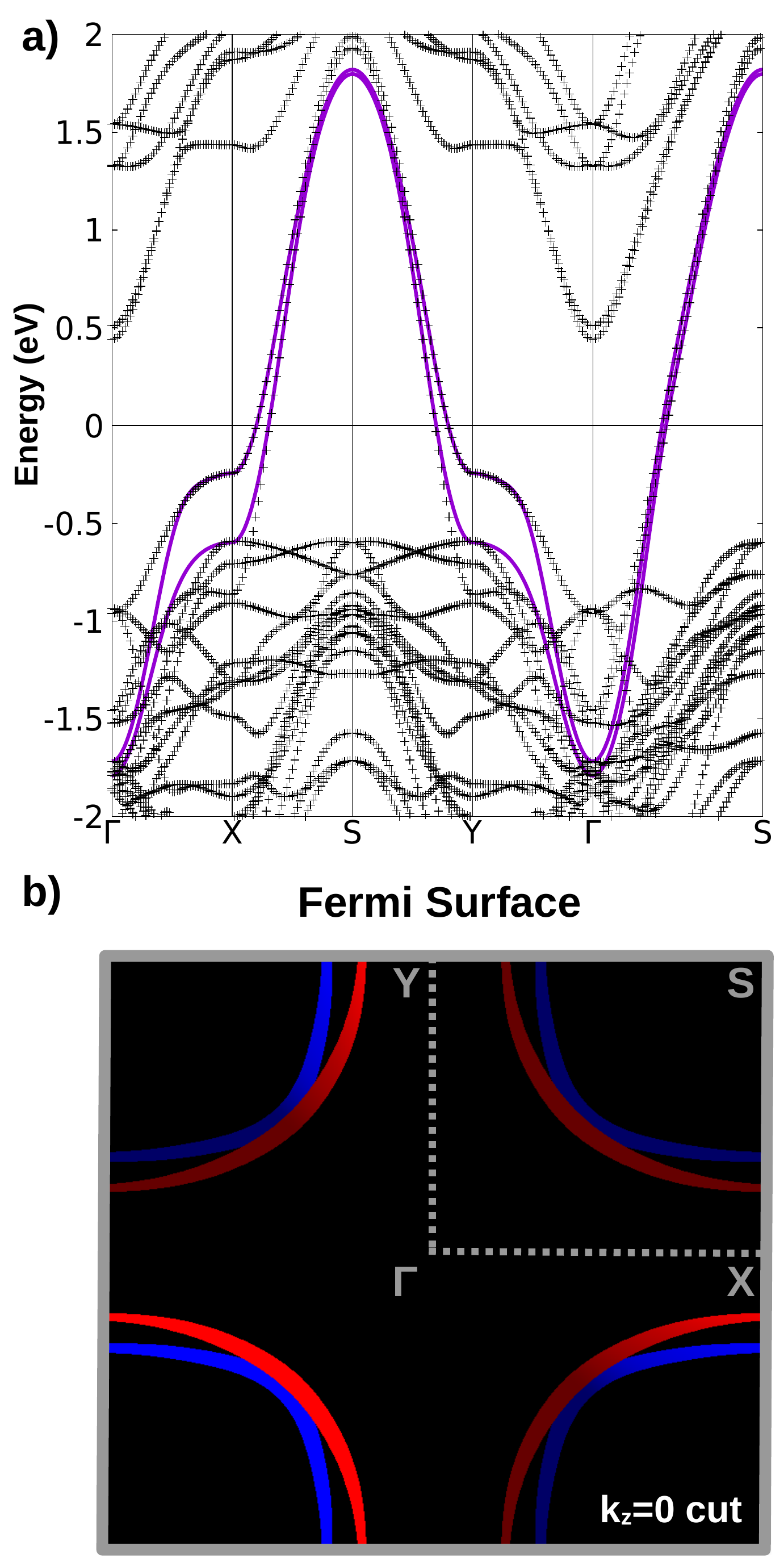}
    \caption{\textbf{a)} All-electron DFT band structure of the YBCO/LMO reference structure (black curves) and Wannier interpolation of the two low-energy CuO$_2$-derived antibonding bands retained in the effective bilayer model (purple curves). The Fermi level is set to zero. \textbf{b)} Corresponding noninteracting Fermi-surface cut in the $k_x$-$k_y$ plane at $k_z=0$. The two Fermi-surface sheets originate from the bonding--antibonding (red--blue) splitting of the CuO$_2$ bilayer.}
    \label{fig:wannier_fs}
\end{figure}

\subsection{Bilayer Hubbard Model}

A major difficulty in describing the YBCO/LMO interface, shown in Fig.~\ref{fig:struct_param} a), lies in the treatment of strong electronic correlations within the cuprate layers. Any low-energy description must account for the competing phases of the cuprate phase diagram, including a Mott insulating state, correlated metallic regimes such as the pseudogap and strange metal, superconductivity phase, and an overdoped Fermi-liquid phase, all evolving as a function of doping in the CuO$_2$ planes. Additional charge, spin, and electronic-nematic orders further enrich this landscape, often leading to intertwined rather than simply competing ordered states~\cite{fradkin_intertwined_orders_2015,comin_damascelli_2016,hayden_tranquada_2024,bourges_loop_currents_2021,zhou_natphys_2025}. A natural framework to address these correlation-driven phenomena is provided by the Hubbard model and its extensions, which have been shown, using a variety of theoretical approaches, to capture salient features of cuprate physics~\cite{lee_nagaosa_wen_2006,qin_hubbard_2022,kancharla_sc_af_2008,senechal_diag_afm_sc,marcello_pg_sc}.

Based on these considerations, we propose [Fig.~\ref{fig:struct_param} b)] the interfacial bilayer CuO$_2$ Hamiltonian $\mathcal{H}_{b}$:
\begin{equation}
    \mathcal{H}_b= \mathcal{H}_{Hub}+\mathcal{H}_{FM},
\end{equation}
where $\mathcal{H}_{Hub}$ is the lattice Hubbard model for a bilayer
\begin{align*}
    \mathcal{H}_{Hub}=&-\sum_{ij\sigma l}t_{ij}^{(l)}[c^{(l)\dagger}_{i\sigma}c^{(l)}_{j\sigma}+h.c.]+\sum_{il}U^{(l)}n^{(l)}_{i\uparrow}n^{(l)}_{i\downarrow}\\
     &-\sum_{i\sigma}t_z[c^{(1)\dagger}_{i\sigma}c^{(2)}_{i\sigma}+h.c.]-\mu\sum_{i\sigma l}n^{(l)}_{i\sigma},
\end{align*}
with $t^{(l)}_{ij}$ the spin-independent on-site and inter-site hopping matrix elements for each pair of sites $(i,j)$ and each layer $l=1,2$. $t_z$ is the interlayer coupling, $U^{(l)}$ the local Coulomb repulsion in each layer, $c^{(l)\dagger}_{i\sigma}$ ($c^{(l)}_{i\sigma}$) the particle creation (annihilation) operator at site $i$ of layer $l$, $n^{(l)}_{i\sigma}=c^{(l)\dagger}_{i\sigma}c^{(l)}_{i\sigma}$ the corresponding number operator, and $\mu$ the chemical potential.

The magnetic proximity effect of the manganite layer is modeled phenomenologically by a local exchange spin-splitting field acting on the Cu sites. At the static mean-field level, we introduce the term
\begin{equation*}
    \mathcal{H}_{FM} = M \sum_{il}c^{(l)\dagger}_{i\sigma}\sigma^z_{\sigma\sigma'}c^{(l)}_{i\sigma'} = M \sum_{il}\left(n^{(l)}_{i\uparrow}-n^{(l)}_{i\downarrow}\right),
\end{equation*}
where $\sigma^z$ is the Pauli matrix associated with spin projections along $\hat z$. The parameter $M$ represents the strength of the local exchange spin-splitting field acting on the Cu states as a consequence of magnetic proximity to the manganite layer. The resulting Cu polarization is determined self-consistently by the interacting solution. Although the coupling is formally Zeeman-like in spin space, $M$ is not a macroscopic applied magnetic field and is not associated with an electromagnetic vector potential. The present Hamiltonian therefore isolates the exchange contribution of the proximity effect: orbital currents, screening, vortex formation, and orbital pair breaking generated by applied or stray magnetic fields are not included.

In this sense, $M$ is a phenomenological control parameter for the strength of the interfacial exchange proximity. Experimentally, the induced Cu polarization depends on the manganite thickness and on the microscopic interface structure~\cite{Dybko_2013,selective_interlayer_coup,Chaudhuri_2023}. In the following, we vary $M$ to track how increasing magnetic proximity affects the coexistence of superconductivity and ferromagnetism in the correlated CuO$_2$ layers.

For convenience, the spin-quantization axis $\hat z$ is chosen along the uniform collinear exchange field. Since the hopping amplitudes and the local Hubbard interaction are spin-rotation invariant, and since neither spin--orbit coupling nor magnetic anisotropy are included, a global unitary rotation in spin space maps any other uniform orientation of the interfacial magnetization onto the form written above. The same Hamiltonian thus describes the spin-sector effect of both in-plane and out-of-plane collinear moments. Their crystallographic orientation becomes physically relevant only when spin--orbit coupling, anisotropic magnetic interactions, or orbital coupling to magnetic fields are retained. This distinction is relevant for cuprate--manganite heterostructures, where experiments report in-plane Cu moments~\cite{selective_interlayer_coup}, as well as a sample-dependent interfacial anisotropy and perpendicular stray-field components~\cite{Uspenskaya_2021,Chaudhuri_2023}.

\subsection{\textit{Ab initio} parametrization}

To determine the noninteracting part of the effective Hubbard model, we use a tetragonal (001)-oriented YBa$_2$Cu$_3$O$_7$ (1 u.c.)/LaMnO$_3$ (1 u.c.) superlattice structural model, hereafter referred to as Y1/L1. LMO is the undoped parent compound of the LSMO and LCMO manganites commonly used in experimental heterostructures and superlattices~\cite{lsmo_pha_diag,CHEN2004295}. In our case the interface has a MnO$_2$--BaO plane termination, as shown in Fig.~\ref{fig:struct_param} a), and we fix the in-plane lattice parameter to $a=3.905$\AA{}, corresponding to epitaxial strain induced by SrTiO$_3$, a substrate commonly used for epitaxial growth. An explicit treatment of LSMO or LCMO would require substantially larger alloy supercells or an additional approximation for the chemical disorder introduced by Sr or Ca substitution. It would also introduce composition-dependent structural and magnetic degrees of freedom, including changes in the MnO$_6$ octahedral rotations and Jahn--Teller distortions. We therefore use the stoichiometric YBCO/LMO heterostructure as a computationally tractable structural reference, rather than as a composition-specific model of the ferromagnetic manganite.

First, we performed a fully relaxed structural DFT+U~\cite{Liechtenstein_1995} calculation using the Vienna Ab initio Simulation Package (VASP)~\cite{vasp1,vasp2,vasp3}, in order to optimize the supercell crystal-structure parameters. The calculations were performed within the Perdew--Burke--Ernzerhof (PBE) generalized gradient approximation (GGA)~\cite{gga}, using projector-augmented-wave (PAW) pseudopotentials~\cite{paw} and a plane-wave energy cutoff of $500$ eV. PAW-PBE pseudopotentials in VASP have been extensively tested for a broad range of materials~\cite{pseudo_test}. Electronic interactions, at the static mean-field level, were considered using a Hubbard term $\textit{U}=5$ eV and a Hund's coupling $\textit{J}=1$ eV for the Mn-3$d$ orbitals, and with $\textit{U}=\textit{J}=0$ eV for Cu-$3d$ states in the same fashion as~\cite{meu_int_dft}. Copper states were initially treated as uncorrelated since the addition of Hubbard-like terms leads to a spurious insulating ground state for Cu-3$d$ states in YBa$_2$Cu$_3$O$_7$ within our DFT+U approximation, as also reported in Ref.~\cite{Damascelli_2008}. The atomic positions and the superlattice size along the $z$ direction were relaxed using a $6\times6\times1$ $k$-point mesh until the forces on each atom were smaller than $0.01$ eV/\AA{}. The relaxed structure [Fig.~\ref{fig:struct_param} a)] converged to an out-of-plane lattice parameter $c=15.51$\AA{}.

From the converged structure, we computed the ground-state electronic structure within the all-electron full-potential approximation (LAPW/APW+lo) as implemented in Wien2K~\cite{wien2k}, using GGA for the exchange-correlation potential, a cutoff parameter $R_{MT}K_{max}=7$, and a $9\times9\times2$ $k$-point mesh. The obtained band structure is shown by the black lines in Fig.~\ref{fig:wannier_fs} a). In an all-electron treatment, core orbitals are explicitly included in the self-consistent cycle, and the full electronic potential is retained, without the use of pseudopotentials. Although computationally more demanding, this approach provides an accurate electronic structure, including improved on-site energies of the atomic orbitals. Since small changes in the electronic dispersion can produce relevant changes in the Fermi surface and low-energy bands, the full-potential treatment is well suited for the subsequent \textit{ab initio} tight-binding parametrization.

The DFT relaxed Y1/L1 structure is used to determine the Cu-derived low-energy dispersion and the hopping parameters of the effective bilayer Hubbard model. The filling of the CuO$_2$ planes is subsequently controlled through the chemical potential, while the ferromagnetic proximity effect produced by doped LSMO or LCMO is introduced independently through the effective exchange field $M$. Thus, varying the chemical potential controls the occupation of the CuO$_2$ low-energy sector and should not be interpreted as simulating Sr or Ca substitution in the manganite. Conversely, neither the ferromagnetic state of the manganite nor its detailed dependence on composition and thickness are derived from the Y1/L1 calculation. The present construction is designed to capture the generic exchange-proximity physics acting on the strongly correlated CuO$_2$ layers, rather than to provide a quantitative, composition-specific description of a particular YBCO/LSMO or YBCO/LCMO heterostructure.

Using a maximally localized Wannier-function basis set~\cite{wannier90}, we then obtained an effective low-energy real-space Hamiltonian by using Wien2wannier~\cite{wien2wannier}. The corresponding band dispersion is shown in purple in Fig.~\ref{fig:wannier_fs} a). These states give rise to the Fermi surface (FS) cut, taken at $k_z=0$, shown in Fig.~\ref{fig:wannier_fs} b), where the two-band coupled bilayer dispersion can be observed. This dispersion represents the main low-energy electronic contribution of the two CuO$_2$ planes at the interface. From the real space Wannier Hamiltonian, we extracted the free-electron tight-binding model parameters [Fig.~\ref{fig:struct_param} b)] listed in Table~\ref{tab:par}.  

\begin{table}[hbt!]
\centering
\begin{tabular}{|c|c|c|c|c|c|c|}
\hline
\textbf{Parameter} & $\mathbf{U}$ & $\mathbf{t}$ & $\mathbf{t'}$ & $\mathbf{t''}$ & $\mathbf{t_z}$ & $\boldsymbol{\varepsilon_d}$ \\
\hline
\hline
\textbf{Values} & 8 & 1 & -0.15 & 0.17 & 0.23 & 0.16 \\
\hline
\end{tabular}
\caption{Model parameters in units of the first-neighbor hopping $|t|\approx0.404$ eV.}
\label{tab:par}
\end{table}

The model parametrization for YBCO and other cuprates has been discussed extensively in previous works~\cite{tb_sacha,tb_weber,arpes_cuprates_review,ele_struct_cuprates_review,downfolding_hubbard_dmrg} using different projection methods. In Table~\ref{tab:r}, we compare the $|t'/t|$ ratio obtained for the YBCO/LMO interface and bulk YBCO with values extracted for YBCO~\cite{tb_sacha,tb_weber}. We find a good agreement for the bulk YBCO in comparison with Ref.~\cite{tb_weber}, which cross-checked the different procedures, including the L\"{o}wdin downfolding.

The appropriate value of the local electron-electron interaction, namely the Hubbard interaction $U$, has also been widely discussed. In this work, we use $U=8t$, a value previously shown to capture the main low-temperature electronic and magnetic phases of cuprates~\cite{senechal_diag_afm_sc}.

\begin{table}[hbt!]
\centering
\begin{tabular}{|c|c|}
\hline
\textbf{Material} & \textbf{$|{t'}/{t}|$} \\
\hline
\hline
Y1/L1 & 0.150 \\
YBCO & 0.149 \\
YBCO~\cite{tb_sacha} & 0.268 \\
YBCO~\cite{tb_weber} & 0.110 \\
\hline
\end{tabular}
\caption{Nearest-neighbor hopping and second nearest-neighbor hopping ratio $|\mathbf{t'}/\mathbf{t}|$ for different cuprate families obtained by projecting \textit{ab initio} electronic structure calculations in localized orbitals.}
\label{tab:r}
\end{table}

Since the two CuO$_2$ planes are geometrically equal in the Y1/L1 structure, we treat layers $(1)$ and $(2)$ as equivalent planes coupled by an interlayer hopping $t_z$ [Fig.~\ref{fig:struct_param} b)], which gives rise to bonding and antibonding states. Finally, as discussed above, the effect of interfacial magnetic ordering is described through the parameter $M$, treated as a tunable interfacial exchange field acting on the spin degrees of freedom.

\subsection{Cluster DMFT}

The bilayer Hamiltonian $\mathcal{H}_{b}$ is solved using cluster dynamical mean-field theory~\cite{kotliar_cdmft_prl,maier_cluster_rmp,kotliar_dmft_rmp_2006,tremblay_lowtemp_2006} (CDMFT). In CDMFT, the lattice problem is mapped onto an effective Anderson impurity problem, in which a finite cluster is embedded in a self-consistent fermionic bath. In the present work, the impurity cluster is a $2\times2$ plaquette within a single CuO$_2$ layer. The two layers of the bilayer are treated as two separate plaquette impurity problems, coupled through the interlayer hopping $t_z$ entering the bilayer CDMFT self-consistency condition, as discussed in Appendix~\ref{bath_par}. In the symmetric Y1/L1 structure considered here, the two layers are equivalent, so the corresponding impurity problems are identical.

For each layer $l$, the Anderson impurity Hamiltonian can be written as
\begin{align*}
    \mathcal{H}^{\rm And}_{l}
    =&\sum_{\alpha\beta\sigma}E^{(l)}_{\alpha\beta\sigma}c^{(l)\dagger}_{\alpha\sigma}c^{(l)}_{\beta\sigma}+U^{(l)}\sum_{\alpha}n^{(l)}_{\alpha\uparrow}n^{(l)}_{\alpha\downarrow}
    \\
    &+\sum_{r\sigma}\varepsilon^{(l)}_{r\sigma}a^{(l)\dagger}_{r\sigma}a^{(l)}_{r\sigma}+\sum_{\alpha r\sigma}\left(\theta^{(l)}_{\alpha r\sigma}c^{(l)\dagger}_{\alpha\sigma}a^{(l)}_{r\sigma}+ h.c.\right)
    \\
    &+\sum_{\alpha r\sigma}\left(\Delta^{(l)}_{\alpha r\sigma}c^{(l)}_{\alpha\sigma}a^{(l)}_{r\bar\sigma}+h.c.\right).
\end{align*}
Here, $\alpha,\beta=1,\ldots,N_c$ label the sites of the $2\times2$ plaquette, with $N_c=4$, while $r$ labels the bath orbitals. The operators $c^{(l)\dagger}_{\alpha\sigma}$ create electrons on the plaquette cluster of layer $l$, whereas $a^{(l)\dagger}_{r\sigma}$ create electrons in the corresponding bath. The matrix $E^{(l)}_{\alpha\beta\sigma}$ contains the one-body terms of the lattice Hamiltonian restricted to the plaquette, including the intralayer hopping amplitudes, the chemical potential, and the local exchange field $M$. The interlayer hopping $t_z$ is not included inside the impurity Hamiltonian itself, but enters the CDMFT self-consistency through the bilayer lattice Green's function.

The normal bath amplitudes $\theta^{(l)}_{\alpha r\sigma}$ describe the hybridization between cluster and bath orbitals with the same spin. The anomalous bath amplitudes $\Delta^{(l)}_{\alpha r\sigma}$ allow the impurity problem to explore superconducting solutions. We stress that $\Delta^{(l)}_{\alpha r\sigma}$ is not an externally imposed pairing field in the lattice Hamiltonian, but rather a variational bath parameter determined self-consistently. Superconductivity is therefore obtained only if the self-consistent solution stabilizes a finite anomalous component.

We employ the Lanczos exact-diagonalization (ED) procedure~\cite{primme} as impurity solver, as implemented in the Python library for quantum cluster methods - pyqcm~\cite{qcm}, which provides the many-body ground state. Our results therefore correspond to zero physical temperature. For the numerical representation of Matsubara Green's functions, we use the fermionic frequency grid $\omega_n=(2n+1)\pi T$, with $T=0.01t$. Here, $T$ is a numerical parameter setting the spacing of the Matsubara-frequency axis and, in this sense, plays a role analogous to an effective temperature, even though the calculation itself is formally performed at $T=0$. Green's functions on the real-frequency axis are obtained by evaluating the same expressions at $z=\omega+i\eta$, with a Lorentzian broadening $\eta=0.1$.

Since ED requires a finite bath, the bath parametrization must be truncated. In our calculations, each $2\times2$ plaquette is coupled to $N_b=8$ bath orbitals. To reduce the number of variational parameters during the CDMFT self-consistency cycle, we impose a simplified $C_2$ symmetry on the bath parametrization~\cite{senechal_diag_afm_sc,senechal_bath_optimization_2010}. This leaves 40 independent bath parameters per plaquette. A detailed description of the bath parametrization and of the bilayer self-consistency is given in Appendix~\ref{bath_par}.

The CDMFT solution gives access to the cluster normal and anomalous Green's functions. In imaginary time, these are defined as
\begin{equation}
    G^{(l)}_{\alpha\sigma,\beta\sigma}(\tau)
    =
    -\left\langle T_\tau c^{(l)}_{\alpha\sigma}(\tau) c^{(l)\dagger}_{\beta\sigma}(0)\right\rangle,
\end{equation}
and
\begin{equation}
    F^{(l)}_{\alpha\sigma,\beta\sigma'}(\tau)
    =
    -\left\langle T_\tau c^{(l)}_{\alpha\sigma}(\tau) c^{(l)}_{\beta\sigma'}(0)\right\rangle,
\end{equation}
where $T_\tau$ is the imaginary-time-ordering operator. Choosing a Nambu basis of the form
\begin{equation}\label{Nambu}
    \Psi_k=\begin{bmatrix} c_{k\uparrow} \\ c^\dagger_{-k\downarrow}\end{bmatrix},
\end{equation}
the total Green's function $\mathbf{G}$ can be written in blocks as
\begin{equation}
    \label{eq:nambu_green}
    \mathbf{G}(k,i\omega_n)=\begin{bmatrix} G(k,i\omega_n) & F(k,i\omega_n) \\ -\overline{F}(k,i\omega_n) & \overline{G}(k,i\omega_n)\end{bmatrix}.
\end{equation}
Here, $\overline{G}$ and $\overline{F}$ denote the corresponding hole-sector components of the Nambu Green's function.

In the collinear magnetic configuration considered here, the normal Green's function is diagonal in spin, while the opposite-spin anomalous sector supports spin-singlet pairing and can also host an $S_z=0$ spin-triplet component when $M\neq0$.

When the anomalous bath amplitudes converge to finite values in the CDMFT cycle, the off-diagonal component of the Nambu Green's function becomes finite, $F(k,i\omega_n)\neq0$, referring to a superconducting state. The superconducting order parameter $\Phi$ can be extracted directly from the anomalous Green's function on the Matsubara-frequency axis:
\begin{equation*}
    \Phi \equiv T\sum_{k,\omega_n}\frac{g(k)}{N_k}F(k,i\omega_n). 
\end{equation*}
Here, $g(k)=\operatorname{sgn}(\cos k_x-\cos k_y)$ is the $d_{x^2-y^2}$ sign form factor. $T$ is the Matsubara-grid parameter introduced above. For a pure even-frequency spin-singlet superconducting state, a global gauge can be chosen such that the anomalous pairing amplitude $F(k,i\omega_n)$ is real on the Matsubara axis, implying that $\Phi$ is real too. However, as we shall see later, upon the application of a magnetic field an odd frequency pairing component develops and the anomalous pairing amplitude becomes complex. In the convention adopted in the calculations~\cite{qcm}, it is convenient to define a real and positive superconducting weight
\begin{equation}\label{ord_sc}
    \langle D\rangle \equiv \sum_{k,\omega_n>0}\frac{1}{N_k}| g(k)F(k,i\omega_n)|.
\end{equation}
Clearly, in the spin-singlet d-wave case realized in the absence of magnetic fields this definition corresponds to a measure of the superconducting order parameter: $2\langle D\rangle=\Phi/T$. The decomposition of $F$ into its even-frequency spin-singlet and odd-frequency spin-triplet components at finite $M$ is not so trivial and will be discussed in detail in Sec.~\ref{subsec:pha_trans}.

\section{\label{sec:res}Results and Discussion}

\begin{figure}
    \centering
    \includegraphics[width=0.9\linewidth]{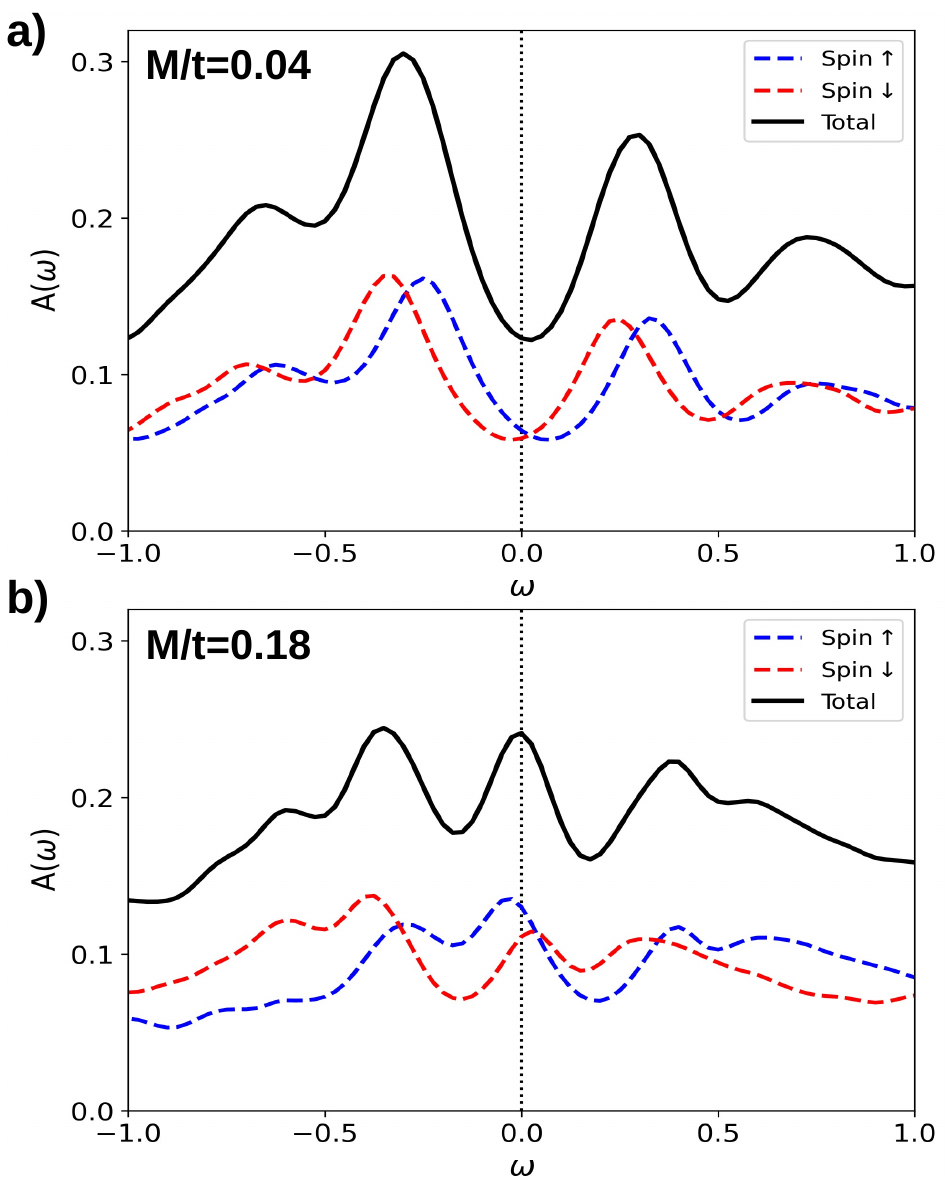}
    \caption{Local spectral functions in the superconducting state at fixed hole doping $p=7\%$, for \textbf{a)} $M/t=0.04$ and \textbf{b)} $M/t=0.18$, the last superconducting solution on the increasing-field branch. Blue and red dashed curves show the spin-up and spin-down contributions, respectively; the black solid curve is their sum. Increasing the exchange field separates the spin-resolved features and brings spectral weight toward the Fermi level, producing the low-energy peak in panel~b). The vertical dotted line marks $\omega=0$. Energies are expressed in units of the nearest-neighbor hopping $|t|\approx0.404$ eV.}
    \label{fig:11}
\end{figure}

We first investigate how the $d$-wave superconducting state is affected by the magnetic moments induced on the Cu sites at the cuprate-manganite interface. All calculations are performed at a fixed hole doping $p=7\%$, corresponding to the occupation of the interfacial CuO$_2$ planes obtained from the \textit{ab initio} calculation of the YBCO/LMO structure [Fig.~\ref{fig:struct_param} a)]. We set the effective interfacial exchange field to $M/t=0.04$, for which the CDMFT solution yields an induced Cu magnetic moment $\langle M^{FM}\rangle\simeq0.02\mu_B$, in accordance with previous experimental~\cite{dead_layer} and DFT+$U$~\cite{meu_int_dft} results.

At $M/t=0$, the converged CDMFT solution is a spin-singlet $d_{x^2-y^2}$ superconductor. Its local spectral function $A(\omega)$ is V-shaped at low energy, and its nearest-neighbor anomalous Green's functions have opposite signs on the $x$ and $y$ bonds~\cite{kancharla_sc_af_2008,senechal_diag_afm_sc,senechal_bath_optimization_2010,civelli_cdmft_sc_2009}. The local and diagonal next-nearest-neighbor anomalous components vanish numerically. The singlet bond amplitude is symmetric under site exchange, $F^s_{12}=F^s_{21}$.

Figure~\ref{fig:11} compares the local spectra at $M/t=0.04$ and $M/t=0.18$. At the lower field, the total density of states retains a pronounced low-energy depletion, while the opposite shifts of the spin-resolved features reveal the exchange splitting. At $M/t=0.18$, the last superconducting solution on the increasing-field branch, spin-split spectral weight has moved toward the Fermi level and a peak appears near $\omega=0$. The superconducting character of this solution is confirmed by its finite anomalous Green's function, discussed below. The splitting is analogous to the Zeeman effect underlying Pauli paramagnetism. For a rigid reference spectrum at fixed chemical potential, the exchange term $+M(n_\uparrow-n_\downarrow)$ would give
\begin{equation}
    A_{\sigma}(\omega) \simeq \frac{1}{2}A_0(\omega-\sigma M), \qquad \sigma=\pm1,
\end{equation}
where $A_0$ is the total spin-degenerate reference spectrum at $M/t=0$. This relation illustrates the direction of the shifts; electronic correlations and self-consistent superconductivity produce a more complex spectral reconstruction in CDMFT. The monolayer calculation in Appendix~\ref{monolayer} shows that exchange splitting and the opposite-spin triplet component persist when the interlayer hopping $t_z$ is removed.

This spin dependence has an important consequence for the anomalous Green's function. Fermionic antisymmetry imposes
\begin{equation}
    F^{\sigma\sigma'}_{ij}(\tau) = -F^{\sigma'\sigma}_{ji}(-\tau).
\end{equation}
For opposite-spin pairing, we introduce the spin-singlet and $S_z=0$ spin-triplet components,
\begin{align}
    F^{s}_{ij}(\tau) &= \frac{1}{2}\left[F^{\uparrow\downarrow}_{ij}(\tau)-F^{\downarrow\uparrow}_{ij}(\tau)\right],\\
    F^{t0}_{ij}(\tau) &= \frac{1}{2}\left[F^{\uparrow\downarrow}_{ij}(\tau)+F^{\downarrow\uparrow}_{ij}(\tau)\right].
\end{align}
At $M/t=0$, the spin-singlet CDMFT solution satisfies
\begin{equation}
    F^{\uparrow\downarrow}_{ij}(\tau) = -F^{\downarrow\uparrow}_{ij}(\tau),
\end{equation}
so that $F^{t0}_{ij}=0$. Since the $d$-wave orbital component is even under exchange of the two spatial coordinates, fermionic antisymmetry requires the spin-singlet pairing function to be even in relative time and, consequently, even in frequency.

\begin{figure}
    \centering
    \vspace{-2mm}
    \includegraphics[width=0.9\linewidth]{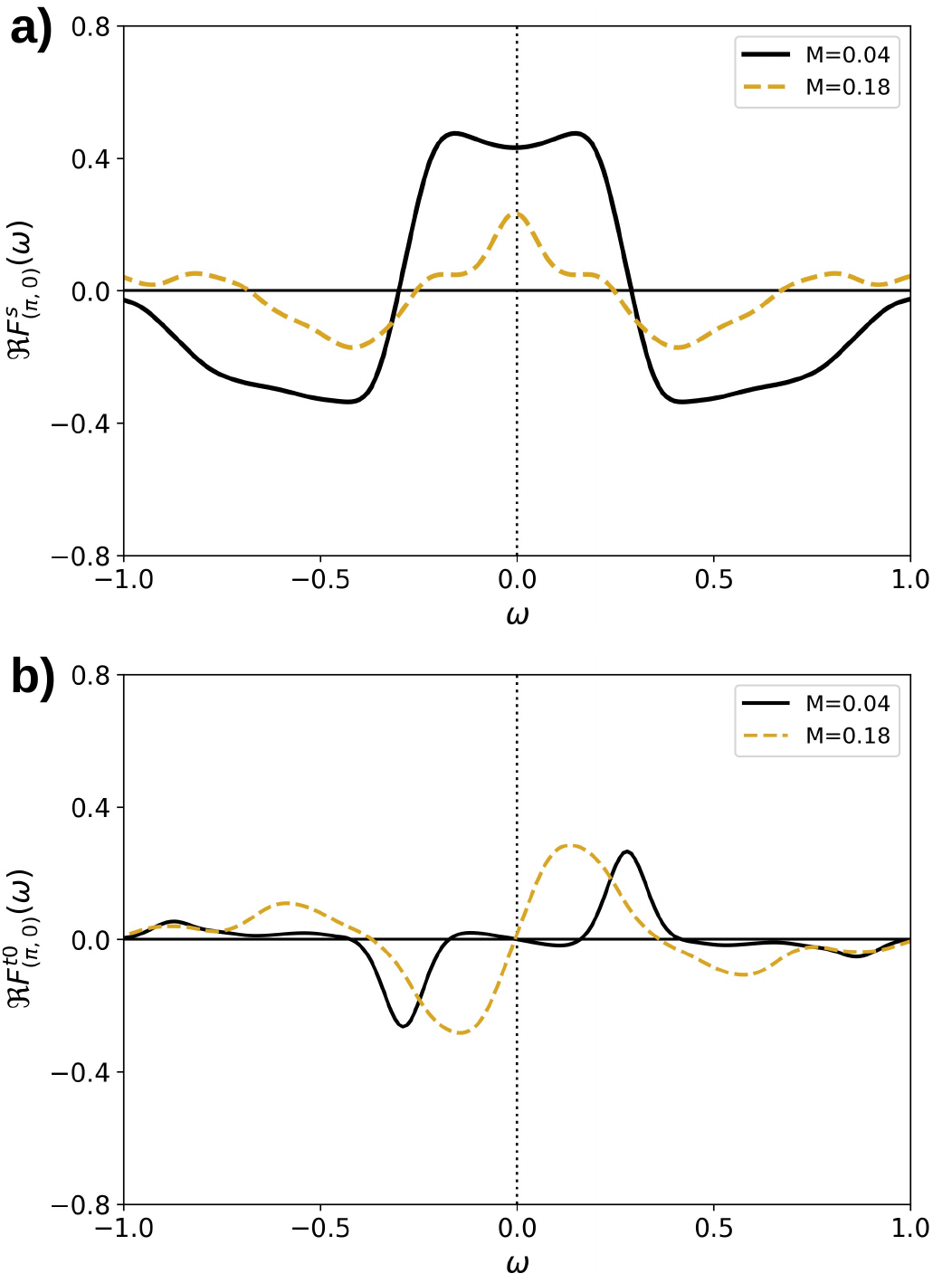}
    \caption{Real parts of the retarded anomalous Green's functions at the antinode $\mathbf{k}=(\pi,0)$: \textbf{a)} even-frequency spin-singlet and \textbf{b)} odd-frequency $S_z=0$ spin-triplet components. Black solid and gold dashed curves correspond to $M/t=0.04$ and $M/t=0.18$, respectively. The singlet real part is even in $\omega$, while the triplet real part is odd. At the higher field, the singlet features are reduced and the opposite-sign triplet lobes move closer to $\omega=0$.}
    \label{fig:re_f_w}
\end{figure}

On the real-frequency axis, exchanging the two time arguments relates retarded and advanced anomalous Green's functions. In equilibrium, and choosing a gauge in which the superconducting amplitude is real, the even-frequency spin-singlet component in the even-parity $d$-wave channel satisfies 
\begin{equation}\label{f1}
    F^{s,R}_{12}(\omega) = \left[F^{s,R}_{12}(-\omega)\right]^*.
\end{equation}
Consequently,
\begin{align}
    \Re F^{s,R}_{12}(\omega) &= \Re F^{s,R}_{12}(-\omega),\\
    \Im F^{s,R}_{12}(\omega) &= -\Im F^{s,R}_{12}(-\omega).
\end{align}
The real part of the even-frequency component is therefore even in $\omega$ [Fig.~\ref{fig:re_f_w} a)], whereas its imaginary part is odd [Fig.~\ref{fig:im_f_w} a)].

At finite $M$, the exchange field makes the spin-up and spin-down propagators inequivalent. As a result,
\begin{equation}
    F^{\uparrow\downarrow}_{ij} \neq -F^{\downarrow\uparrow}_{ij},
\end{equation}
and a finite $S_z=0$ spin-triplet component is generated. Since the Hamiltonian contains no spin-flip processes, spin projection along the magnetization axis remains conserved, so pairing mechanisms continue to involve opposite spins, and no equal-spin triplet component is produced.

The induced $S_z=0$ triplet component is symmetric under spin exchange and our solution shows that it belongs to the same even-parity $d$-wave orbital channel as the singlet component. Fermionic antisymmetry, therefore, requires this triplet contribution to be odd in relative time, or equivalently odd in frequency. On the real-frequency axis it satisfies
\begin{equation}\label{f2}
    F^{t0,R}_{12}(\omega) = -\left[F^{t0,R}_{12}(-\omega)\right]^*.
\end{equation}
It follows that
\begin{align}
    \Re F^{t0,R}_{12}(\omega) &= -\Re F^{t0,R}_{12}(-\omega),\\
    \Im F^{t0,R}_{12}(\omega) &= \Im F^{t0,R}_{12}(-\omega).
\end{align}
Thus, the real part of the odd-frequency component is odd in $\omega$ [Fig.~\ref{fig:re_f_w} b)], whereas its imaginary part is even [Fig.~\ref{fig:im_f_w} b)].

The finite-$M$ superconducting solution thus contains an even-frequency spin-singlet component and an induced odd-frequency $S_z=0$ spin-triplet component in the same spatial $d$-wave channel. Indeed, Figure~\ref{fig:re_f_w} shows their real parts at the antinode $\mathbf{k}=(\pi,0)$ for the two fields used in Fig.~\ref{fig:11}. The singlet curve in panel~a) is symmetric under $\omega\rightarrow-\omega$, whereas the triplet curve in panel~b) is antisymmetric, as required by Eqs.~\eqref{f1} and~\eqref{f2}. Increasing $M/t$ from $0.04$ to $0.18$ reduces the singlet features and moves the opposite-sign triplet lobes toward the Fermi level. The imaginary parts on the real-frequency axis and the corresponding Matsubara functions are presented in Appendix~\ref{f_odd}. We now quantify the field evolution of these pairing components before examining their relation to the magnetic response.

This mixed-frequency structure is a symmetry-allowed consequence of Zeeman-split singlet superconductivity and can be already reproduced by the BCS-like quadratic $d$-wave model developed in Appendix~\ref{bcs}. That reference calculation will allow us to distinguish the response expected at the BCS level from the correlation-driven enhancement found below.

\subsection{Evolution of superconductivity under the exchange field}\label{subsec:pha_trans}

\begin{figure}[ht!]
    \centering
    \includegraphics[width=0.9\linewidth]{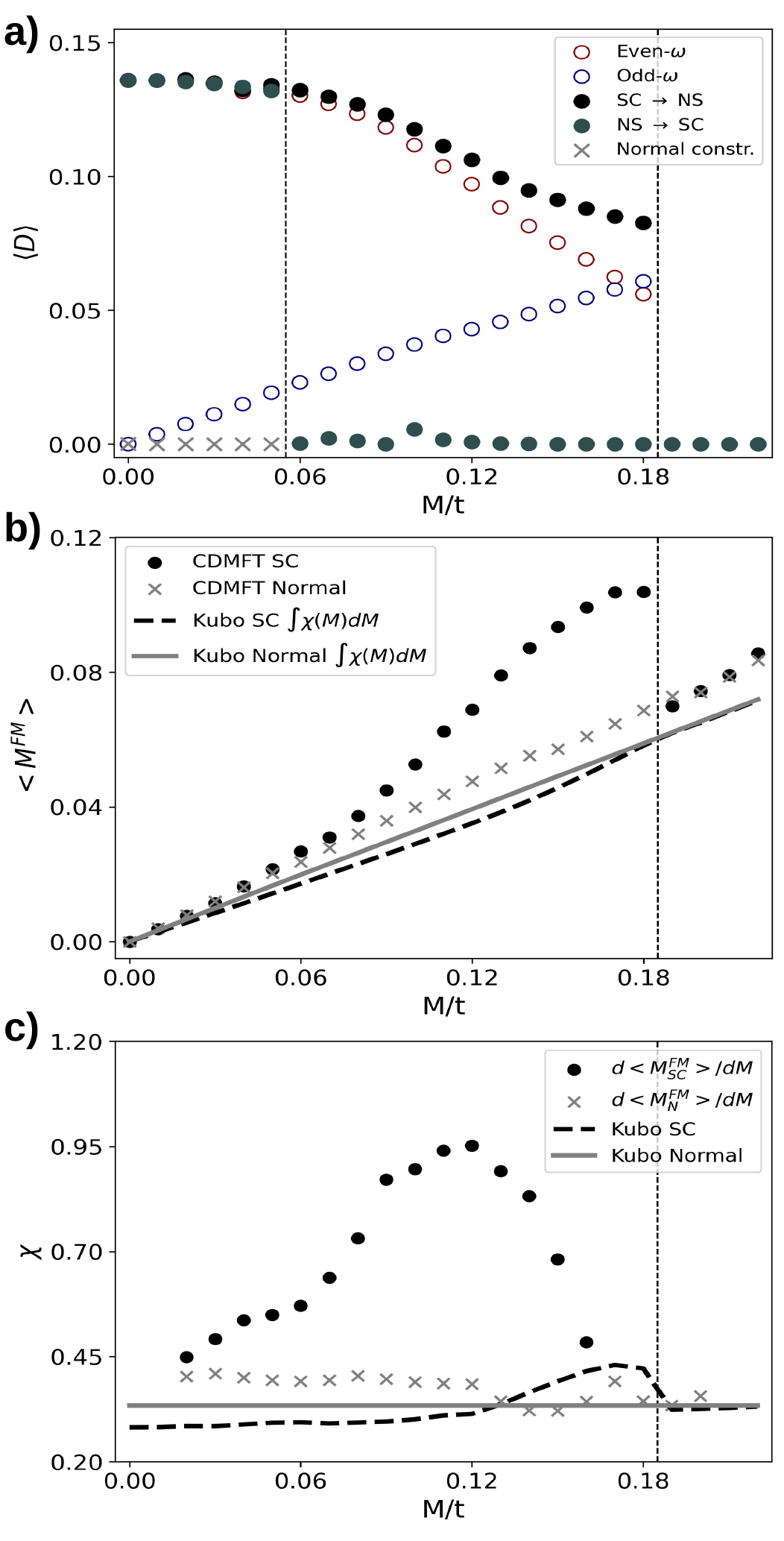}
    \caption{Field evolution at fixed hole doping $p=7\%$. \textbf{a)} Superconducting amplitude $\langle D\rangle$: black filled circles follow increasing $M/t$ from $0$ to $0.22$ (SC $\to$ NS), and dark green filled circles follow the reverse scan (NS $\to$ SC). Their different switching fields reveal hysteresis. Open red and blue circles are $D_{\mathrm{even}}$ and $D_{\mathrm{odd}}$ on the superconducting increasing-field branch; gray crosses denote constrained normal solutions. The two vertical dashed lines bracket the transition fields of the two scans. \textbf{b)} Induced magnetic moment: black circles and gray crosses are the direct CDMFT results along the increasing-field and constrained normal branches, respectively. Black dashed and gray solid lines are the magnetization values reconstructed by integrating the corresponding fixed-density bubbles. \textbf{c)} Fixed-density susceptibilities: black circles and gray crosses are numerical derivatives of the direct CDMFT magnetization values; black dashed and gray solid lines are the superconducting and normal bubble contributions. In panels~b) and~c), the vertical dashed line marks the loss of superconductivity on increasing $M$.}
    \label{fig:ord_evo_m}
\end{figure}

We now investigate the evolution of the superconducting state upon increasing the interfacial exchange field $M$, keeping the hole doping fixed at $p=7\%$. As shown by the black filled circles in Fig.~\ref{fig:ord_evo_m} a), superconductivity (SC) persists up to $M/t\simeq0.18$, beyond which the anomalous weight $\langle D\rangle$ introduced in Eq.~\eqref{ord_sc} is abruptly suppressed and the system jumps to a normal-state (NS) solution. Calculations initialized  in the NS at $M/t=0.22$ and continued toward lower fields remain normal over part of the field interval where the increasing-field branch is superconducting; they recover SC only at a lower field, as shown by the jump from  the dark green to the black circles. This hysteresis supports a first-order transition. The gray crosses in Fig.~\ref{fig:ord_evo_m} a) show a constrained normal-state branch continued to $M/t=0$ by setting the anomalous bath components to zero in the CDMFT self-consistency.

The corresponding critical exchange-field scale is comparable to the amplitude of the $d$-wave superconducting gap, consistently with a Zeeman-driven Pauli pair-breaking mechanism, $M_c\sim\Delta$. This scale is also consistent with the redistribution of spectral weight toward the Fermi level in Fig.~\ref{fig:11} b). The quadratic BCS reference in Appendix~\ref{bcs} illustrates the associated Zeeman spectral evolution.

To characterize separately the evolution of the spin-singlet even-frequency and the spin-triplet odd-frequency anomalous components, the opposite-spin anomalous Green's function can be decomposed on the Matsubara-frequency axis as
\begin{equation}
    F^{\uparrow\downarrow}(k,i\omega_n) = F^s(k,i\omega_n)+F^{t0}(k,i\omega_n).
\end{equation}
In the gauge adopted here, the even-frequency spin-singlet component is purely real, whereas the odd-frequency spin-triplet component is purely imaginary, as shown explicitly in Appendix~\ref{f_odd} (Fig.~\ref{fig:f_iwn}), so we decompose $F^{\uparrow\downarrow}$ directly as
\begin{equation}\label{eq:F_even_odd}
    \begin{aligned}
        F^{\uparrow\downarrow}(k,i\omega_n) &= \Re F^s(k,i\omega_n)+i\Im F^{t0}(k,i\omega_n)\\
        &\equiv F_{\mathrm{even}}(k,i\omega_n)+iF_{\mathrm{odd}}(k,i\omega_n).
    \end{aligned}
\end{equation}
with $F_{\mathrm{even}}(k,i\omega_n)\equiv\Re F^s(k,i\omega_n)$ and $F_{\mathrm{odd}}(k,i\omega_n)\equiv\Im F^{t0}(k,i\omega_n)$.

Substituting Eq.~\eqref{eq:F_even_odd} into Eq.~\eqref{ord_sc} allows us to express the integrated anomalous weight as
\begin{align}
    \langle D\rangle &= \sum_{k,\omega_n>0}\frac{1}{N_k}\left|F_{\mathrm{even}}(k,i\omega_n)+iF_{\mathrm{odd}}(k,i\omega_n)\right|\nonumber\\
    &=\sum_{k,\omega_n>0}\frac{1}{N_k}\sqrt{F_{\mathrm{even}}^2(k,i\omega_n)+F_{\mathrm{odd}}^2(k,i\omega_n)}.
\end{align}
To quantify the even- and odd-frequency contributions to the total superconducting amplitude, we define
\begin{align}
    D_{\mathrm{even}} &= \sum_{k,\omega_n>0}\frac{1}{N_k}\left|F_{\mathrm{even}}(k,i\omega_n)\right|,\\
    D_{\mathrm{odd}} &= \sum_{k,\omega_n>0}\frac{1}{N_k}\left|F_{\mathrm{odd}}(k,i\omega_n)\right|.
\end{align}
Their evolution is shown by the open red and blue circles, respectively, in Fig.~\ref{fig:ord_evo_m} a). The even-frequency singlet weight decreases with increasing $M$, while the odd-frequency triplet weight grows and becomes comparable to the singlet contribution at $M/t\simeq0.18$, immediately before superconductivity collapses on the increasing-field branch. The exchange field therefore suppresses the overall superconducting amplitude while increasing the relative importance of the induced triplet component.

\subsection{Superconductivity-enhanced magnetic polarization}\label{subsec:magnetic_polarization}

We next examine the induced magnetic moment,
\begin{equation}\label{eq:magnetic_moment}
    \langle M^{FM}\rangle = \left|\langle n_\uparrow\rangle-\langle n_\downarrow\rangle\right|.
\end{equation}
Fig.~\ref{fig:ord_evo_m} b) compares the direct CDMFT magnetization values along the increasing-field branch (black circles) and the constrained normal branch (gray crosses), at the same hole density and exchange field. Within the superconducting regime, the black circles lie above the gray crosses: superconductivity enhances the induced magnetic polarization. The separation is particularly pronounced in the upper part of the superconducting branch. Once superconductivity collapses, the increasing-field solution joins the normal branch. The black dashed and gray solid curves are reconstructed from the bubble susceptibilities and will be discussed in Sec.~\ref{subsec:beyond_bubble}.

This enhancement contrasts with the usual competition between ferromagnetism and conventional singlet superconductivity, for which pairing is expected to reduce the magnetic polarization, as illustrated by the quadratic BCS reference in Appendix~\ref{bcs}. For instance, local magnetic moments were found to be suppressed in the predominantly singlet $d_{x^2-y^2}$ superconducting phase of cuprate thin films subjected to high magnetic fields~\cite{ybco_M}. The induced odd-frequency spin-triplet component might suggest a possible explanation for the enhanced response. Such field-induced correlations can arise from a parent spin-singlet state, as also demonstrated for Zeeman-coupled transition-metal dichalcogenide superconductors~\cite{sano2026}. Their presence alone, however, does not explain the CDMFT magnetization enhancement. Indeed, the BCS calculations in Appendix~\ref{bcs} (Fig.~\ref{fig:bcs_norm}) contain the same mixed-frequency structure, but the superconducting magnetization remains below its normal-state counterpart throughout the field interval shown.

\subsection{Magnetic susceptibility beyond the one-particle bubble}\label{subsec:beyond_bubble}

To determine how much of the magnetization enhancement is captured by the one-particle propagators, we compare the direct CDMFT susceptibility with the bare Kubo bubble. Along each self-consistent branch at fixed hole density, the direct response is
\begin{equation}\label{eq:chi_cdmft_fixed_density}
    \chi_{p}^{\mathrm{CDMFT}} = \left.\frac{d\langle M^{FM}\rangle}{dM}\right|_p.
\end{equation}
These numerical derivatives are shown by black circles for the increasing-field branch and gray crosses for the constrained normal branch in Fig.~\ref{fig:ord_evo_m} c).

The bubble is constructed from the same interacting Green's functions, so it retains their one-particle correlation effects but omits vertex corrections and the self-consistent response of the self-energy to the field. In our Nambu basis, the bare magnetic vertex is the identity, $\mathbf{\Sigma}_0=\mathbf{1} $, and the fixed-chemical-potential bubble is
\begin{equation}
    \chi_{\mu}^{\mathrm{bub}} = -T\sum_{k,\omega_n}\frac{1}{N_k}\operatorname{Tr}\left[\mathbf{\Sigma}_0\mathbf{G}(k,i\omega_n)\mathbf{\Sigma}_0\mathbf{G}(k,i\omega_n)\right].
\end{equation}
Evaluating the trace yields
\begin{equation}
    \chi_{\mu}^{\mathrm{bub}} = -T\sum_{k,\omega_n}\frac{1}{N_k}\left[G^2+\overline{G}^{\,2}-2F\overline{F}\right].
\end{equation}
Here, $\overline{G}$ and $\overline{F}$ are the hole-sector components defined in Eq.~\eqref{eq:nambu_green}.

Substituting the even and odd-frequency components defined in Eq.~\eqref{eq:F_even_odd}, we obtain
\begin{equation}
    \chi_{\mu}^{\mathrm{bub}}=-T\sum_{k,\omega_n}\frac{1}{N_k}\left[G^2+\overline{G}^{\,2}+2\left(F_{\mathrm{even}}^2-F_{\mathrm{odd}}^2\right)\right].
\end{equation}
The susceptibility can therefore be decomposed as
\begin{equation}
    \chi_{\mu}^{\mathrm{bub}} = \chi^{GG}+\chi^{FF}_{\mathrm{even}}+\chi^{FF}_{\mathrm{odd}},
\end{equation}
where
\begin{align}\label{eq:chi_decomp}
    \chi^{GG} &= -T\sum_{k,\omega_n}\frac{G^2+\overline{G}^{\,2}}{N_k},\\
    \chi^{FF}_{\mathrm{even}} &= -2T\sum_{k,\omega_n}\frac{F_{\mathrm{even}}^2}{N_k},\\
    \chi^{FF}_{\mathrm{odd}} &= +2T\sum_{k,\omega_n}\frac{F_{\mathrm{odd}}^2}{N_k}.
\end{align}
All three sums run over the full Matsubara-frequency grid, consistently with the trace expression and Appendix~\ref{bcs}. In the solutions considered here, $\chi^{GG}>0$, while $\chi^{FF}_{\mathrm{even}}<0$ and $\chi^{FF}_{\mathrm{odd}}>0$: the even-frequency singlet term suppresses the bubble response and the odd-frequency triplet term enhances it. The same sign structure is already present in the quadratic BCS reference. To compare with the fixed-density CDMFT derivatives, we use the corresponding fixed-density bubble $\chi_p^{\mathrm{bub}}$; the conversion and its small numerical correction are detailed in Appendix~\ref{bcs} (Eq.~\eqref{eq:fixed_density_bubble_main}).

In Fig.~\ref{fig:ord_evo_m} c), the constrained normal bubble (gray solid line) is nearly field independent and Pauli-like. The superconducting bubble (black dashed line) starts below it and develops an upturn at higher fields, eventually exceeding the normal bubble near the end of the superconducting branch. This Zeeman-driven enhancement is already captured by the quadratic BCS model in Appendix~\ref{bcs}: as the exchange field approaches the gap scale, spin-split gap-edge spectral weight moves toward the Fermi level and increases the low-energy density of states. The same spectral redistribution is visible in Fig.~\ref{fig:11} b). In the quadratic reference, the positive odd-frequency contribution $\chi_{\mathrm{odd}}$ of Eq.~\eqref{eq:chi_cdmft_fixed_density} also grows as $M$ approaches the gap scale and reinforces the high-field upturn of the total susceptibility; related paramagnetic responses have been reported for odd-frequency pairs in other settings~\cite{higashitani2013}.

In the CDMFT calculations, the direct response is in fact much larger than this one-particle approximation. The superconducting derivatives in Fig.~\ref{fig:ord_evo_m} c) display a broad maximum well above the bubble before decreasing toward the transition. The direct normal response also exceeds its bubble over most of the field range, but the additional enhancement is much stronger in the superconducting branch. This difference identifies a correlated response beyond the dressed one-particle bubble, involving vertex corrections and self-consistent feedback. The positive odd-frequency contribution to the bubble alone is therefore insufficient to account for the superconductivity-enhanced magnetization.

To connect this susceptibility comparison back to the magnetization in Fig.~\ref{fig:ord_evo_m} b), we integrate the fixed-density bubble along each branch:
\begin{equation}\label{eq:integrated_bubble_moment}
    \langle M^{FM}\rangle_{\mathrm{bub}}(M) = \int_0^M \chi_p^{\mathrm{bub}}(M')\,dM'.
\end{equation}
The resulting black dashed and gray solid curves correspond to the superconducting and constrained normal branches, respectively. Although the superconducting bubble susceptibility exceeds the normal bubble at high fields, its integrated magnetization remains below the normal bubble result over the superconducting field range, as in the quadratic $d$-wave BCS reference in Appendix~\ref{bcs} [Fig.~\ref{fig:bcs_norm} b)]. Both integrated curves underestimate the direct CDMFT magnetization, especially in the superconducting state. The bubble thus fails to reproduce the central result, $\langle M^{FM}\rangle_{\mathrm{SC}}>\langle M^{FM}\rangle_{\mathrm{N}}$: superconductivity amplifies a correlation contribution beyond the one-particle response.

\subsection{\label{subsec:normal_state} Spin-polarized pseudogap state at strong exchange field}

\begin{figure*}[]
    \centering
    \includegraphics[width=0.85\linewidth]{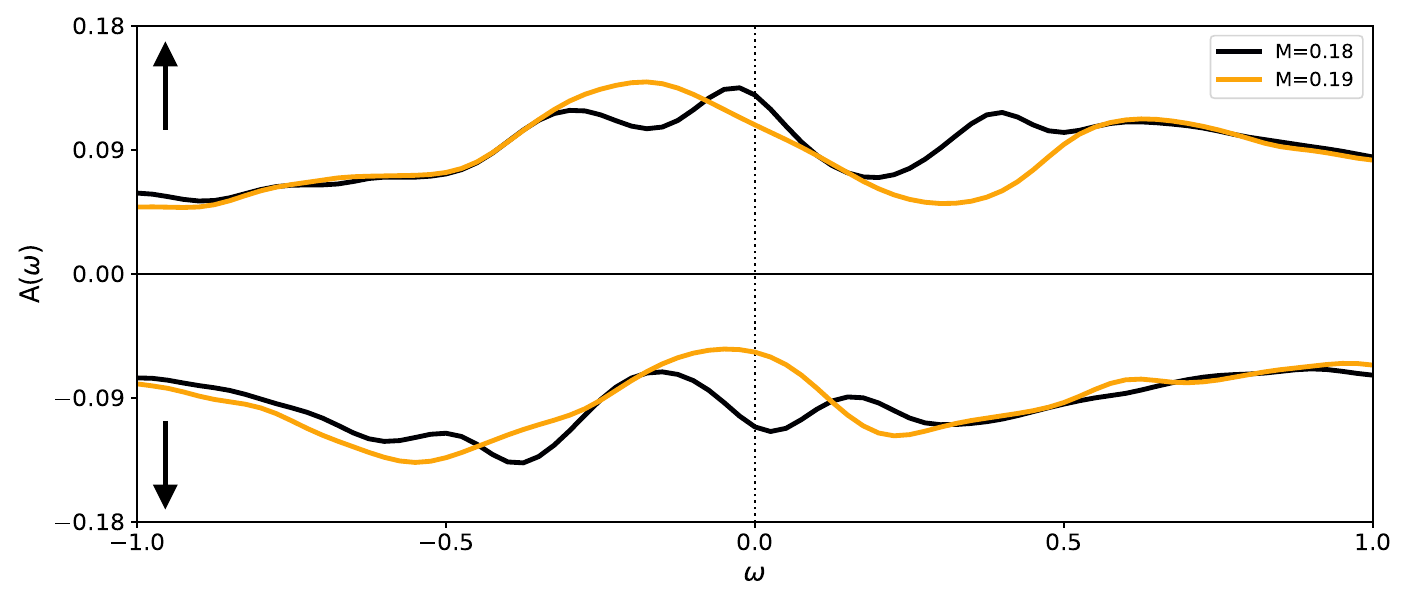}
    \caption{Spin-resolved spectral functions immediately before and after the superconducting-to-normal state transition, at $M/t=0.18$ (black) and $M/t=0.19$ (orange), respectively. For clarity, the spin-up spectral function $A_\uparrow(\omega)$ is plotted with positive sign, whereas the spin-down spectral function $A_\downarrow(\omega)$ is plotted with opposite sign.}
    \label{fig:M_evo_dos}
\end{figure*}

For $M/t>0.18$, the superconducting pairing amplitude vanishes and the system enters a normal state. A marked reconstruction of the frequency-dependent spectral function $A(\omega)$ is visible in Fig.~\ref{fig:M_evo_dos}. 
Although superconductivity has disappeared, a depletion of low-energy spectral weight persists near $\omega=0$. This depletion is particularly pronounced in the spin-down channel. The resulting state therefore exhibits pseudogap-like spectral properties analogous to those of the normal state above $T_c$ in the underdoped region of the cuprate phase diagram. In the present case, however, the pseudogap is strongly spin polarized by the interfacial exchange field.

\begin{figure*}[]
    \centering
    \includegraphics[width=0.8\linewidth]{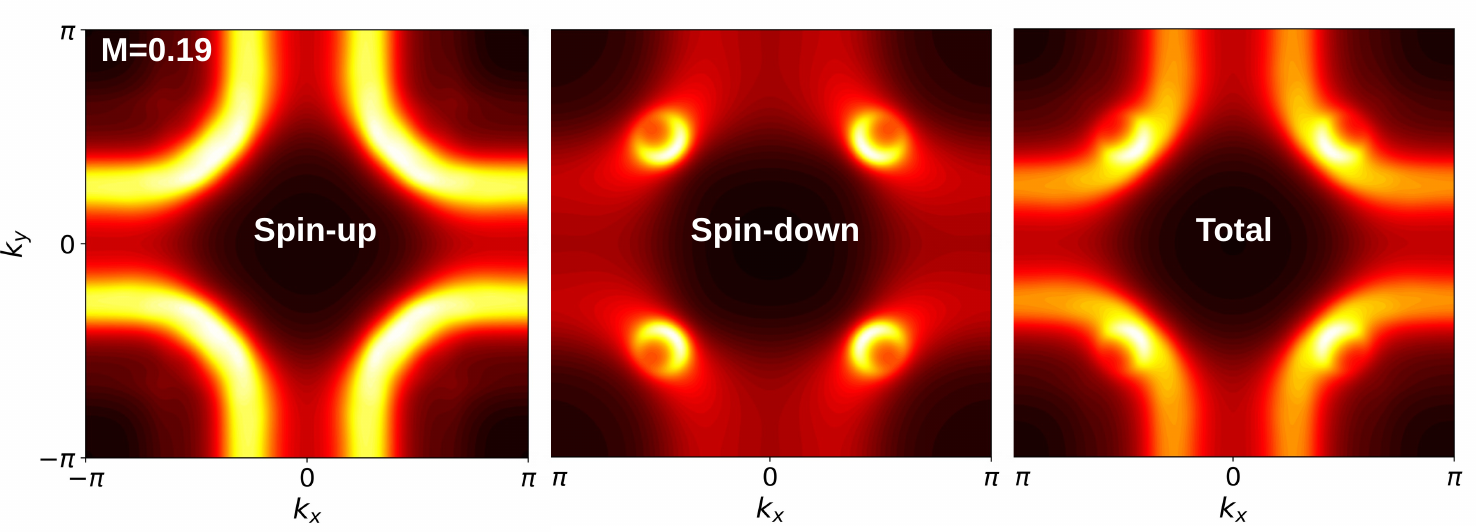}
    \caption{Momentum-resolved spectral weight $A(\mathbf{k},\omega=0)$ in the spin-polarized normal state at $M/t=0.19$, immediately beyond the superconducting transition. The panels show the spin-up contribution (left), spin-down contribution (center), and their sum (right).}
    \label{fig:M19_fs}
\end{figure*}

This spin-selective character is more clearly displayed in Fig.~\ref{fig:M19_fs}, where we show the spin-resolved and total momentum-resolved spectral function $A(\mathbf{k},\omega\rightarrow0)$ for $M/t=0.19$ at the Fermi level, corresponding to the golden curves in Fig.~\ref{fig:M_evo_dos}. The spin-up projection shown in the left panel retains an extended and nearly continuous Fermi contour, with only moderate differentiation between the nodal region $\mathbf{k}\sim(\pi/2,\pi/2)$ and the antinodal regions $\mathbf{k}\sim(\pi,0)$ and $(0,\pi)$. In sharp contrast, the spin-down spectrum in the central panel displays a strong suppression of antinodal spectral weight. The remaining low-energy weight is confined to short segments around the nodal directions, producing a Fermi-arc-like structure. When the two spin components are added, the total spectral function shown in the right panel retains a clear Fermi-arc-like pattern, reminiscent of that observed by ARPES in several cuprate compounds above $T_c$~\cite{arpes_cuprates_review}. In the present case, however, the two spin contributions are strongly differentiated by the exchange-induced Cu magnetic moments at the interface. A detailed dependence of the spin-resolved spectral function cuts as a function of the exchange field $M$ is shown in Appendix~\ref{fs_evo}.

In Appendix~\ref{pseudogap} we analyze both spectral function and self-energy low-energy frequency dependence at $\mathbf{k}\sim(\pi,0)$, where a clear presence of the characteristic structure of the strongly correlated normal-state pseudogap found in our previous calculations~\cite{sakai_s_wave_pseudogap_2013} can be observed. This confirms that the low-energy depletion is not simply a rigid Zeeman displacement of the superconducting gap, but a genuine correlation-driven pseudogap phenomenon. We therefore identify the resulting phase as a spin-polarized pseudogap state: a correlated normal state in which superconductivity has been suppressed by the interfacial exchange field and a pseudogap emerges with a strongly spin-selective texture. Despite the abrupt loss of the anomalous solution, the spectral evolution suggests that the high-field superconducting state and the polarized pseudogap share the same underlying correlated background, but it is worth highlighting that this observation does not, by itself, establish thermodynamic continuity between the two solutions.

\subsection{\label{subsec:local_moment}Correlated local moments and superconductivity-enhanced polarization}

The spin-polarized pseudogap normal state provides a useful starting point for interpreting the anomalously large superconducting magnetization. To distinguish the formation of an instantaneous moment from its net polarization, we evaluate
\begin{equation}\label{eq:main_mloc}
    m_{\mathrm{loc}}^2 \equiv \left\langle\left(n_{i\uparrow}-n_{i\downarrow}\right)^2\right\rangle = \langle n_i\rangle-2d_{\mathrm{occ}},
\end{equation}
where $d_{\mathrm{occ}}=\langle n_{i\uparrow}n_{i\downarrow}\rangle$ is the on-site double occupancy. This equal-time quantity measures the instantaneous local-moment weight. Our CDMFT results in the superconducting phase indicate $m_{\mathrm{loc}}^2\simeq0.84$, approximately constant for different $M$ values, whereas in the BCS quadratic model described in Appendix~\ref{bcs} ($U=0$ one-particle tight-binding model), we obtained $m_{\mathrm{loc}}^2\simeq0.50$. Strong correlations, therefore, generate a large instantaneous local-moment weight by suppressing double occupancy due to the high-energy cost. Within the coexistence regime, however, $m_{\mathrm{loc}}^2$ changes by less than one percent between the superconducting and normal phases and is, in fact, slightly smaller in the superconducting solution (approximately $0.836$), than in the constrained normal state solution (approximately $0.84$).

After superconductivity collapses, the spin-polarized pseudogap retains essentially the same large local-moment weight, $m_{\mathrm{loc}}^2\simeq0.84$. Since CDMFT maps the lattice problem onto a self-consistent cluster Anderson impurity model, in the strong-$U$ regime, this admits the familiar Kondo-like language of moment formation and dynamical screening by the bath. The pseudogap is, therefore, naturally viewed as a regime of well-developed but still dynamical correlation-induced moments, rather than as a lattice of immobile classical spins. The combined behavior of $m_{\mathrm{loc}}^2$ and $\langle M^{FM}\rangle$ is consistent with superconductivity reorganizing the dynamical compensation of these pre-existing moments, leaving a larger fraction of the fermionic weight polarized by the exchange field. This screening-based picture provides an intuitive interpretation of the numerical result, although establishing the underlying dynamical mechanism would require an explicit two-particle analysis beyond the scope of the present work.

\section{\label{sec:conc}Conclusions and Outlook}

In summary, inspired by the competition of ferromagnetism and superconductivity realized in manganite-cuprate interfaces, we used CDMFT to investigate a bilayer Hubbard model that isolates the exchange-proximity effect acting on the first CuO$_2$ planes of the interface. Increasing the effective interfacial exchange field $M$ drives the system from a ferromagnetic $d$-wave superconducting state with an induced Cu spin polarization to a correlated normal state. In the superconducting regime, the collinear exchange field also generates the symmetry-allowed odd-frequency $S_z=0$ spin-triplet component that accompanies Zeeman-split singlet pairing. At a critical value of $M$, the anomalous solution disappears abruptly and the low-energy spectrum reconstructs, signalizing a first order transition into a strongly spin-selective, correlation-driven pseudogap.

The central result is that, throughout the coexistence regime, the superconducting solution carries a larger Cu magnetic moment than the corresponding constrained normal solution at the same density and exchange field. The fixed-density bare magnetic Kubo bubble captures the negative even-frequency anomalous contribution, the positive odd-frequency contribution, and the susceptibility upturn associated with Zeeman-shifted gap-edge spectral weight. It anyways strongly underestimates the direct CDMFT response. The missing response is therefore a correlation-driven contribution beyond the dressed one-particle bubble, encoded in vertex and collective feedback, and it is substantially stronger in the superconducting branch than in the normal state solution.

The instantaneous local moment clarifies the character of this enhancement. Strong correlations increase $m_{\mathrm{loc}}^2$ from approximately $0.49$--$0.50$ in the uncorrelated model to approximately $0.84$ in CDMFT. Superconductivity leaves this large local-moment weight essentially unchanged, while increasing the static polarization. The anomalous magnetization therefore reflects a stronger polarization of already formed correlated moments rather than the formation of moments with a larger instantaneous amplitude. The persistence of the same local-moment scale in the spin-polarized pseudogap further supports the view that the high-field superconducting state develops from this correlated background.

The predicted magnetic response can be connected directly to element and layer-sensitive probes already used in closely related heterostructures and superlattices. X-ray magnetic circular dichroism (XMCD) at the Cu and Mn $L_{2,3}$ edges has been used to probe induced Cu moments and their coupling to the manganite magnetization in both YBCO/LCMO and YBCO/LSMO multilayers~\cite{dead_layer,Sen_2016,Prajapat_2018}. Measurements across $T_c$ on a series of samples with different manganite thicknesses could therefore search for a Cu-specific enhancement, or a change of slope, of the dichroic signal in the superconducting state while simultaneously monitoring the Mn response. Polarized neutron reflectometry could provide complementary information on the depth profile of the magnetization and has already revealed large superconductivity-induced rearrangements in cuprate--manganite superlattices~\cite{Hoppler_2009,Prajapat_2018}. The $^{63}$Cu Knight shift is, in principle, an additional probe of the Cu spin susceptibility, as established in YBCO~\cite{Takigawa_1989_knight}. However, for a single ultrathin interface, the small volume fraction of interfacial Cu sites and the finite applied field required for nuclear magnetic resonance (NMR), together with the associated orbital and vortex response, would complicate a quantitative comparison. Such a measurement may become more realistic in superlattices containing many equivalent interfaces, where the interfacial contribution is amplified.

The calculated spin-resolved spectral function $A_{\sigma}(\mathbf{k},\omega)$ makes spin-resolved and angle-resolved photoemission spectroscopy (ARPES) a particularly direct test of our results. In an experimental geometry in which the ultrathin cuprate layer terminates the sample surface, in-situ ARPES could probe the interfacial electronic structure without the usual obstruction associated with a buried interface. ARPES has already been used to resolve the low-energy electronic structure of YBCO surfaces~\cite{Hossain_2008_arpes}, while in-situ measurements on epitaxial LSCO thin films demonstrate the feasibility of applying the technique to artificially grown cuprate layers~\cite{Kim_2018_arpes,Zhong_2022_arpes}. Spin-resolved ARPES has also been performed in cuprate superconductors~\cite{Iwasawa_2023_spin_arpes}. Applied to a surface-accessible cuprate--manganite heterostructure, it could test the predicted spin-dependent reconstruction of the superconducting spectrum, the accumulation of low-energy spectral weight close to the collapse of superconductivity, and the emergence of a spin-selective pseudogap beyond it. For buried interfaces, spin-polarized cross-sectional scanning tunneling spectroscopy (STS) would provide a complementary local probe and non-polarized STM/STS measurements had already demonstrated access to the electronic structure of YBCO/LCMO interfaces~\cite{Chien2013}. A correlated observation of the Cu-moment anomaly and the spin-resolved spectral reconstruction would provide a stringent experimental test of the correlation-enhanced magnetic response and of the spin-polarized pseudogap predicted here.

Overall, our results show that strong electronic correlations can reverse the conventional magnetic response of spin-singlet superconductivity. Instead of suppressing the interfacial polarization, the superconducting solution can polarize already formed correlated moments more strongly than the underlying normal state. Cuprate--manganite interfaces therefore provide a promising platform for investigating how unconventional superconductivity, local-moment physics, and the pseudogap reshape magnetic proximity beyond a one-particle description.

\begin{acknowledgments}

The authors acknowledge financial support from the Brazilian funding agencies CNPq (Grants No.~402919/2021-1, 201149/2024-9, and 305726/2023-4) and CAPES, as well as from the Université Paris-Saclay Programme de Financement des Cotutelles Internationales de Doctorat. W.~H.~B. acknowledges financial support from the Air Force Office of Scientific Research (AFOSR) under Grant No.~FA9550-24-1-0279. W.~H.~B. and V.~A.~M.~L. thank E.~Miranda and M.~Radovic for useful discussions. M.~C. and V.~A.~M.~L. acknowledge support from D.~S\'en\'echal with the implementation of the calculations using the \texttt{pyqcm} package, and insightful discussions with M.~Aprili and M.C. Aguiar. We also acknowledge the computational resources provided by the National Laboratory for Scientific Computing (LNCC/MCTI, Brazil) through the SDumont supercomputer (\url{http://sdumont.lncc.br}), and by the Moulon M\'esocentre Informatique (Universit\'e Paris-Saclay) through access to the Ruche supercomputer.

\end{acknowledgments}

\appendix
\section{Bath parameters and bilayer self-consistency}
\label{bath_par}

In CDMFT with exact diagonalization (CDMFT-ED), each CuO$_2$ layer is described by a $2\times2$ plaquette impurity coupled to a finite set of bath orbitals. The two plaquette impurity problems are coupled at the level of the bilayer CDMFT self-consistency through the interlayer hopping $t_z$. In the symmetric Y1/L1 structure considered in this work, the two layers are equivalent, and the two impurity problems have identical bath parametrizations.

In the following, we adopt the notation and bath parametrization conventions used in the pyqcm implementation of CDMFT-ED~\cite{qcm}. For a given layer $l$, the impurity Green's function is written in Nambu space as a $2N_c\times2N_c$ matrix,
\begin{equation}
    \mathbf{G}^{(l)}_{\rm imp}(z)
    =
    \left[z\mathbf{1}-\mathbf{E}^{(l)}-\mathbf{\Gamma}^{(l)}(z)-\mathbf{\Sigma}^{(l)}(z)\right]^{-1},
\end{equation}
where $z$ denotes a complex frequency, $\mathbf{E}^{(l)}$ is the one-body cluster matrix, $\mathbf{\Gamma}^{(l)}(z)$ is the hybridization function generated by the bath, and $\mathbf{\Sigma}^{(l)}(z)$ is the cluster self-energy. The self-energy is obtained from the Dyson equation
\begin{equation}
    \mathbf{\Sigma}^{(l)}(z)
    =
    \mathbf{G}^{(l)-1}_{0,\rm imp}(z)-\mathbf{G}^{(l)-1}_{\rm imp}(z),
\end{equation}
where $\mathbf{G}^{(l)}_{0,\rm imp}$ is the noninteracting impurity Green's function.

The bilayer nature of the problem enters through the cluster-projected lattice Green's function,
\begin{equation}
    \mathbf{G}_{\rm loc}(z)
    =
    \frac{N_c}{N}\sum_{\tilde{\mathbf{k}}}\left[z\mathbf{1}-\mathbf{T}(\tilde{\mathbf{k}})-\mathbf{\Sigma}(z)\right]^{-1}.
\end{equation}
Here, the sum runs over the reduced Brillouin zone associated with the $2\times2$ cluster tiling of each CuO$_2$ plane. The bilayer hopping matrix has the block form
\begin{equation}
    \mathbf{T}(\tilde{\mathbf{k}})
    =
    \begin{pmatrix}
    \mathbf{T}^{(1)}(\tilde{\mathbf{k}}) & \mathbf{T}_z \\
    \mathbf{T}_z^\dagger & \mathbf{T}^{(2)}(\tilde{\mathbf{k}})
    \end{pmatrix},
\end{equation}
where $\mathbf{T}^{(1)}$ and $\mathbf{T}^{(2)}$ contain the intralayer hopping terms, chemical potential and exchange field for each layer, while $\mathbf{T}_z$ contains the interlayer hopping $t_z$ between the two CuO$_2$ planes. In contrast, the self-energy is taken to be local to each plaquette impurity and block diagonal in the layer index,
\begin{equation}
    \mathbf{\Sigma}(z)
    =
    \begin{pmatrix}
    \mathbf{\Sigma}^{(1)}(z) & 0 \\
    0 & \mathbf{\Sigma}^{(2)}(z)
    \end{pmatrix}.
\end{equation}
The CDMFT self-consistency condition is imposed by matching each impurity Green's function to the corresponding layer block of the bilayer local Green's function,
\begin{equation}
    \mathbf{G}^{(l)}_{\rm imp}(z)
    =
    \left[
    \mathbf{G}_{\rm loc}(z)
    \right]_{ll}.
\end{equation}
In practice, this condition is implemented by optimizing the bath parameters entering $\mathbf{\Gamma}^{(l)}(z)$ at each CDMFT iteration.

The hybridization function is generated by the finite bath according to
\begin{equation}
    \mathbf{\Gamma}^{(l)}(z)
    =
    \boldsymbol{\theta}^{(l)}\left[z\mathbf{1}-\boldsymbol{\varepsilon}^{(l)}\right]^{-1}\boldsymbol{\theta}^{(l)\dagger}.
\end{equation}
Here, $\boldsymbol{\varepsilon}^{(l)}$ is the bath-energy matrix, while $\boldsymbol{\theta}^{(l)}$ contains both the normal and anomalous bath-cluster hybridization amplitudes. In the following, we omit the layer index $l$ for simplicity.

Using $N_c=4$ and $N_b=8$, the bath-energy matrix $\boldsymbol{\varepsilon}$ is a $2N_b\times2N_b$ matrix. We choose it to be diagonal in the bath-orbital basis,
\begin{equation*}
    \boldsymbol{\varepsilon}
    =
    \begin{pmatrix}
        \boldsymbol{\varepsilon}_{\uparrow} & 0 \\
        0 & \boldsymbol{\varepsilon}_{\downarrow}
    \end{pmatrix},
\end{equation*}
with
\begin{equation*}
    \boldsymbol{\varepsilon}_{\sigma}
    =
    \begin{pmatrix}
        \varepsilon^{\sigma}_1 & 0 & \cdots & 0 \\
        0 & \varepsilon^{\sigma}_2 & \cdots & 0 \\
        \vdots & \vdots & \ddots & \vdots \\
        0 & 0 & \cdots & \varepsilon^{\sigma}_8
    \end{pmatrix}.
\end{equation*}
This parametrization contains 16 bath-energy parameters. We further impose $\varepsilon_i^\uparrow=\varepsilon_i^\downarrow$, reducing this number to 8.

The bath-cluster hybridization matrix has the block structure
\begin{equation*}
    \boldsymbol{\theta}
    =
    \begin{pmatrix}
        \boldsymbol{\theta}_{\uparrow} & \boldsymbol{\Delta}_{\uparrow\downarrow} \\
        \boldsymbol{\Delta}_{\uparrow\downarrow}^{\dagger} & \boldsymbol{\theta}_{\downarrow}
    \end{pmatrix}.
\end{equation*}
The blocks $\boldsymbol{\theta}_{\sigma}$ describe normal hybridization processes between cluster and bath orbitals with the same spin, while $\boldsymbol{\Delta}_{\uparrow\downarrow}$ couples pair creation/annihilation (anomalous) amplitudes between cluster and bath, allowing superconducting solutions to emerge self-consistently. To reduce the number of independent bath parameters, we impose a simplified $C_2$ symmetry on the bath parametrization~\cite{senechal_diag_afm_sc,senechal_bath_optimization_2010}. This symmetry relates the hybridization amplitudes associated with equivalent sites of the $2\times2$ plaquette, while still allowing for the magnetic and superconducting solutions considered in this work. The sign structure in the following matrices encodes this symmetry constraint and allows the bath to support the relevant plaquette symmetry channels, including the $d$-wave superconducting component.

The same-spin impurity-bath hopping blocks, with $\sigma=\uparrow,\downarrow$, are parametrized as
\begin{equation*}
    \boldsymbol{\theta}_{\sigma}
    =
    \begin{pmatrix}
        \theta^{\sigma}_1 & \theta^{\sigma}_2 & \theta^{\sigma}_3 & \theta^{\sigma}_4 & \theta^{\sigma}_5 & \theta^{\sigma}_6 & \theta^{\sigma}_7 & \theta^{\sigma}_8 \\
        \theta^{\prime\sigma}_1 & \theta^{\prime\sigma}_2 & -\theta^{\prime\sigma}_3 & -\theta^{\prime\sigma}_4 & \theta^{\prime\sigma}_5 & \theta^{\prime\sigma}_6 & -\theta^{\prime\sigma}_7 & -\theta^{\prime\sigma}_8 \\
        \theta^{\sigma}_1 & -\theta^{\sigma}_2 & \theta^{\sigma}_3 & -\theta^{\sigma}_4 & \theta^{\sigma}_5 & -\theta^{\sigma}_6 & \theta^{\sigma}_7 & -\theta^{\sigma}_8 \\
        \theta^{\prime\sigma}_1 & -\theta^{\prime\sigma}_2 & -\theta^{\prime\sigma}_3 & \theta^{\prime\sigma}_4 & \theta^{\prime\sigma}_5 & -\theta^{\prime\sigma}_6 & -\theta^{\prime\sigma}_7 & \theta^{\prime\sigma}_8
    \end{pmatrix},
\end{equation*}
while the anomalous bath block is parametrized equivalently as
\begin{equation*}
    \boldsymbol{\Delta}_{\uparrow\downarrow}
    =
    \begin{pmatrix}
        \Delta_1 & \Delta_2 & \Delta_3 & \Delta_4 & \Delta_5 & \Delta_6 & \Delta_7 & \Delta_8 \\
        \Delta^\prime_1 & \Delta^\prime_2 & -\Delta^\prime_3 & -\Delta^\prime_4 & \Delta^\prime_5 & \Delta^\prime_6 & -\Delta^\prime_7 & -\Delta^\prime_8 \\
        \Delta_1 & -\Delta_2 & \Delta_3 & -\Delta_4 & \Delta_5 & -\Delta_6 & \Delta_7 & -\Delta_8 \\
        \Delta^\prime_1 & -\Delta^\prime_2 & -\Delta^\prime_3 & \Delta^\prime_4 & \Delta^\prime_5 & -\Delta^\prime_6 & -\Delta^\prime_7 & \Delta^\prime_8
    \end{pmatrix}.
\end{equation*}

\begin{figure}
    \centering
    \includegraphics[width=0.95\linewidth]{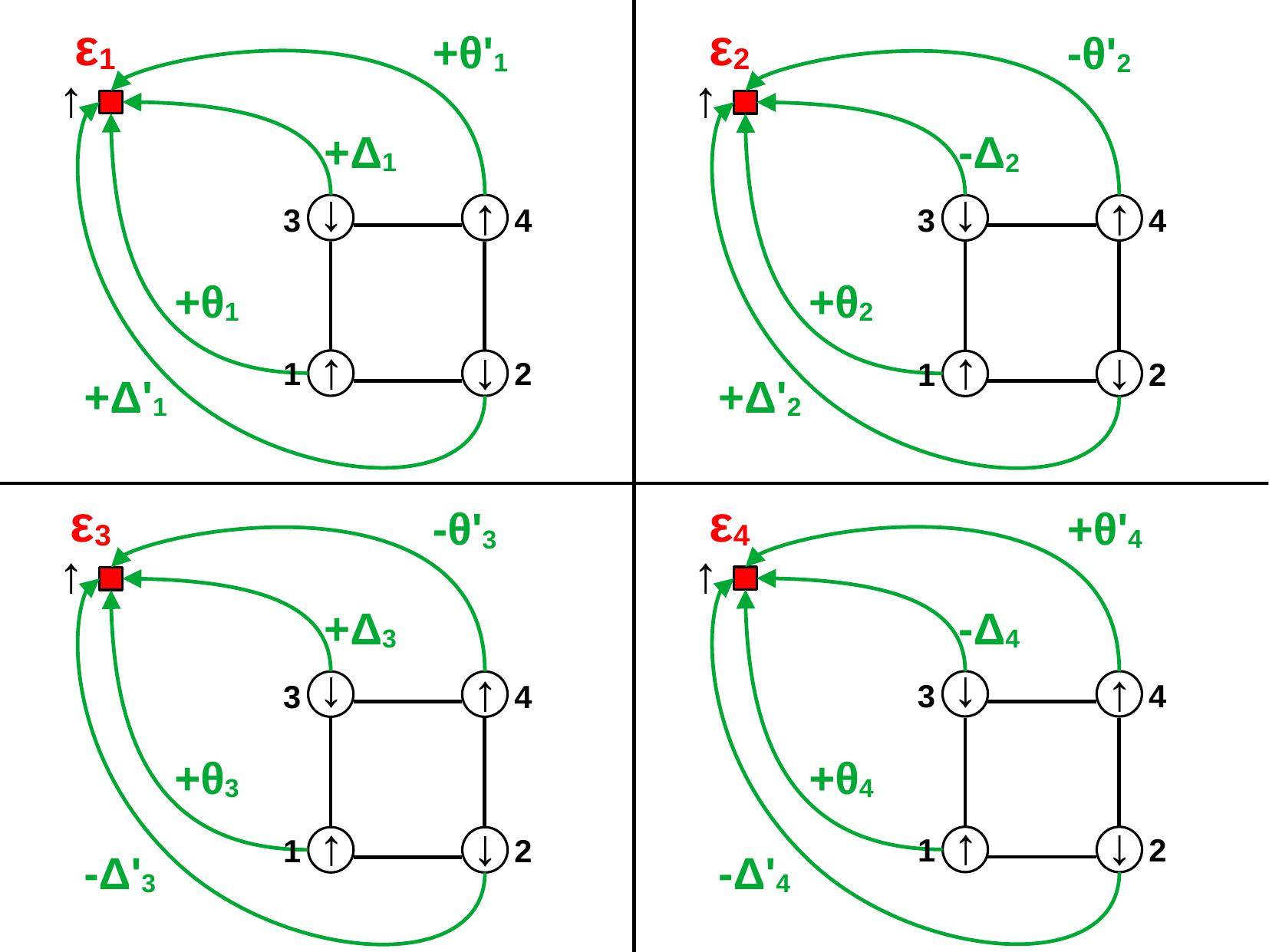}
    \caption{Bath parametrization of a $2\times2$ plaquette with $N_b=4$ bath orbitals and simplified $C_2$ symmetry. The $N_b=8$ parametrization follows the same construction, and the opposite-spin sector is equivalent.}
    \label{fig:c2}
\end{figure}

There are, in principle, 32 parameters associated with $\boldsymbol{\theta}$ and 32 additional parameters associated with $\boldsymbol{\Delta}$. We reduce this parameter space by taking the anomalous bath amplitudes $\Delta_i$ and $\Delta_i^\prime$ to be real, and by imposing $\theta_i^\uparrow=\theta_i^\downarrow$ and $\theta_i^{\prime\uparrow}=\theta_i^{\prime\downarrow}$. Together with the simplified $C_2$ symmetry encoded in the matrices above, this leaves 16 independent parameters in $\boldsymbol{\theta}$, 16 in $\boldsymbol{\Delta}$, and 8 in $\boldsymbol{\varepsilon}$, for a total of 40 bath parameters per plaquette. Since the two layers are equivalent, the same parametrization is used for both impurity problems. Figure~\ref{fig:c2} illustrates the construction for one spin channel and $N_b=4$ bath orbitals.

\section{Monolayer model}\label{monolayer}

\begin{figure}
    \centering
    \includegraphics[width=1\linewidth]{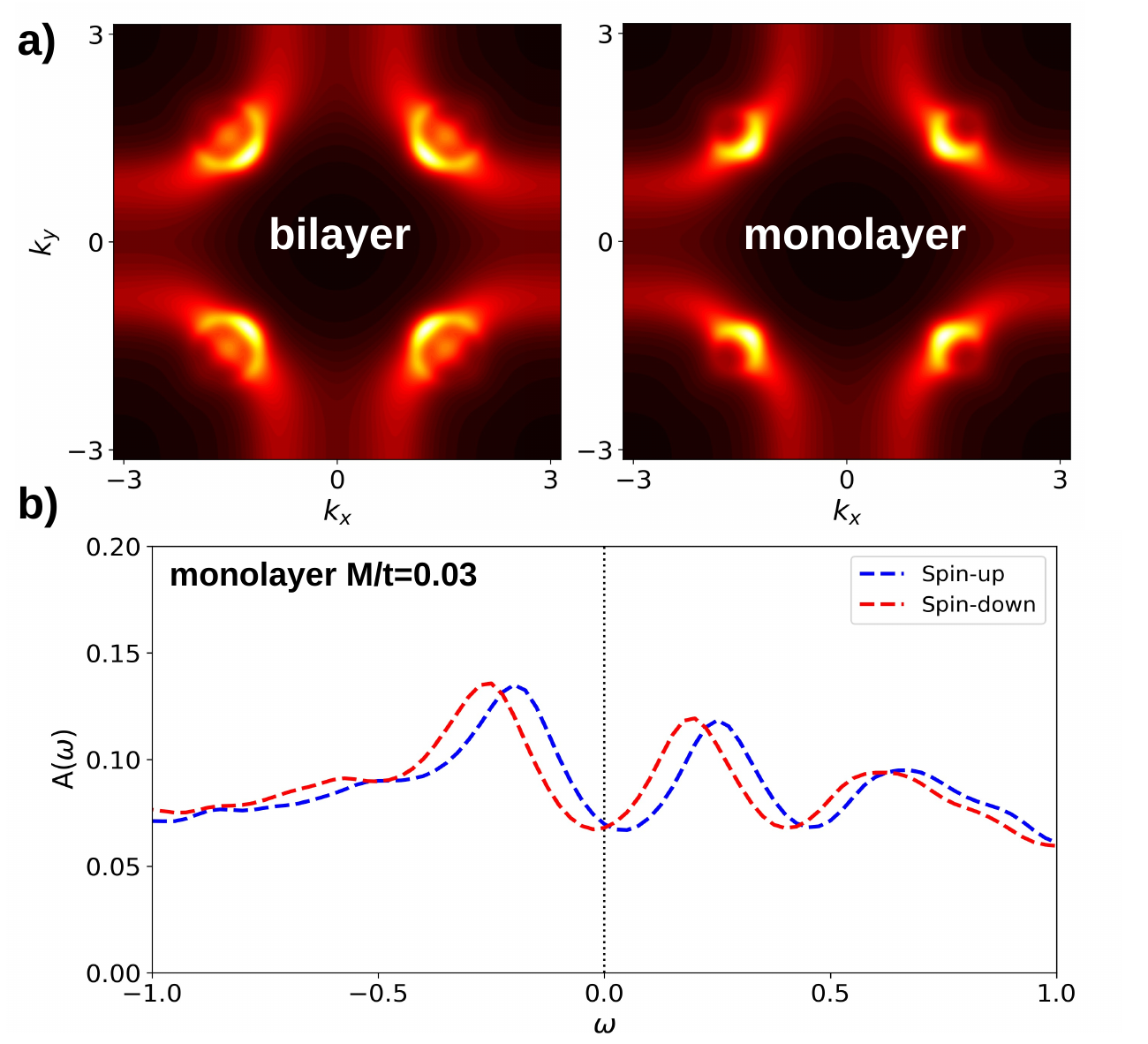}
    \caption{Monolayer CDMFT results at fixed hole doping $p=7\%$. \textbf{a)} Momentum-resolved spectral function $A(\mathbf{k},\omega=0)$ at $M/t=0$ for the bilayer model (left) and for the monolayer obtained by setting $t_z=0$ (right). \textbf{b)} Spin-resolved local spectral functions of the superconducting monolayer at $M/t=0.03$. The spin-up channel is indicated in blue and the spin-down channel in red.}
    \label{fig:mono}
\end{figure}

To isolate the role of interlayer coupling, we also performed calculations for a monolayer CuO$_2$ model, separating effects intrinsic to the bilayer structure from those generated locally by the exchange field. The monolayer calculations were performed by setting the interlayer hopping parameter $t_z$ to zero. Fig.~\ref{fig:mono} a) compares the CDMFT spectral function cut at the Fermi level $A(\mathbf{k},\omega=0)$ of the bilayer (left panel) and monolayer (right panel) models at $M/t=0$. In both cases the presence of the $d$-wave state results in a spectral weight concentrated around the nodal directions $|k_x|=|k_y|$. The bilayer displays two main features associated with its bonding and antibonding components, while the monolayer model has only one contribution. This assures that the two-pocket pattern is a consequence of $t_z$, rather than a reconstruction caused by the exchange field. In Fig.~\ref{fig:mono} b) we present the CDMFT spin-resolved spectral function for the monolayer at $M/t=0.03$, corresponding to $\langle M^{FM}\rangle\simeq0.02\mu_B$ and $p=7\%$ in the one-band model. The converged solution indicates a clear spin-splitting contribution around the Fermi energy, demonstrating that, even at $t_z=0$, the exchange field breaks the spin symmetry of the system for finite $M$.

\section{Complementary anomalous spectral}\label{f_odd}

\begin{figure}
    \centering
    \includegraphics[width=0.68\linewidth]{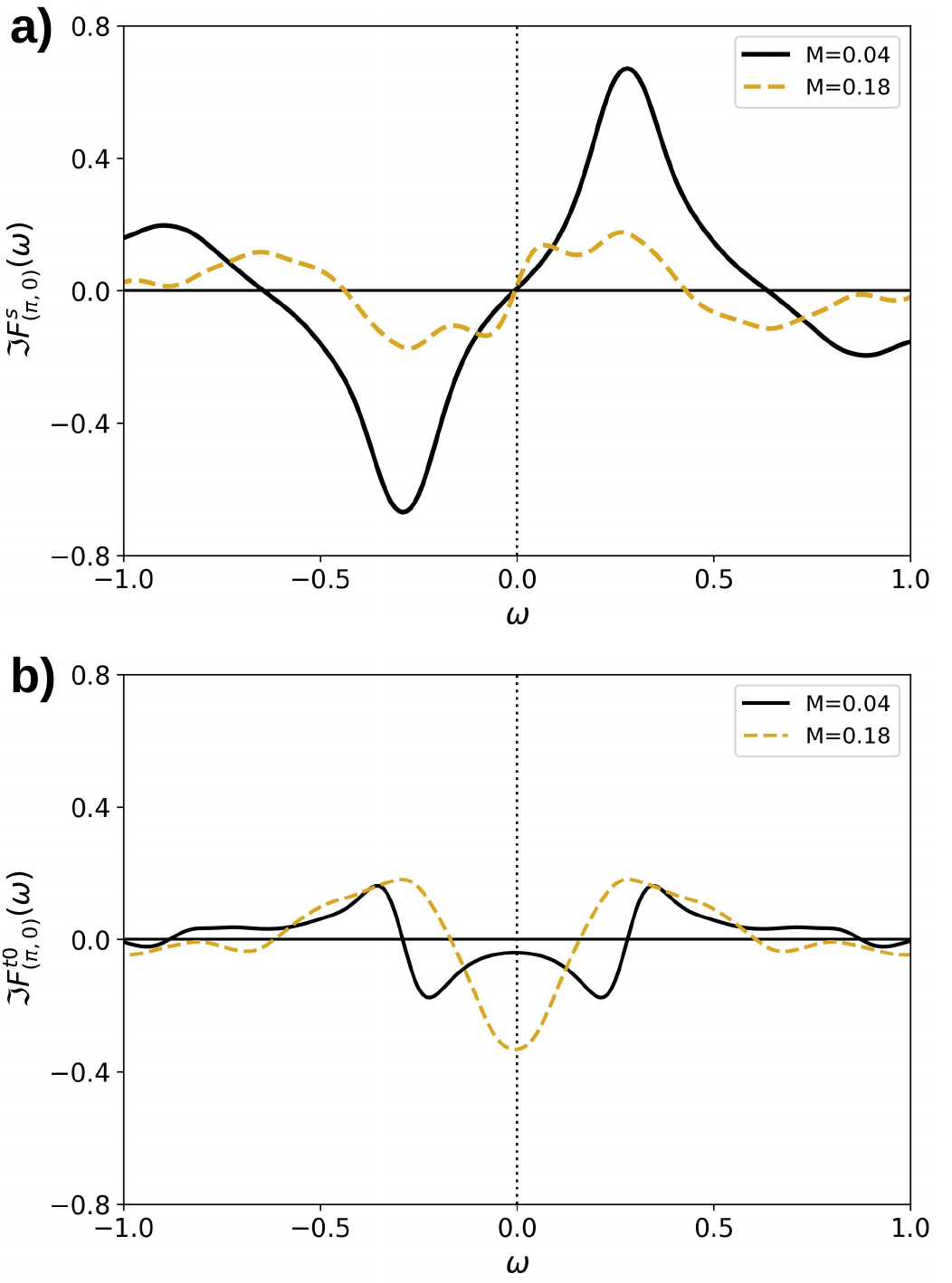}
    \caption{Imaginary part of the anomalous Green's function \textbf{a)} even-frequency spin-singlet component and \textbf{b)} odd-frequency $S_z=0$ spin-triplet component at the antinode $\mathbf{k}=(\pi,0)$ for $M/t=0.04$ (full black lines) and $M/t=0.18$ (golden dashed lines).}
    \label{fig:im_f_w}

    \includegraphics[width=0.68\linewidth]{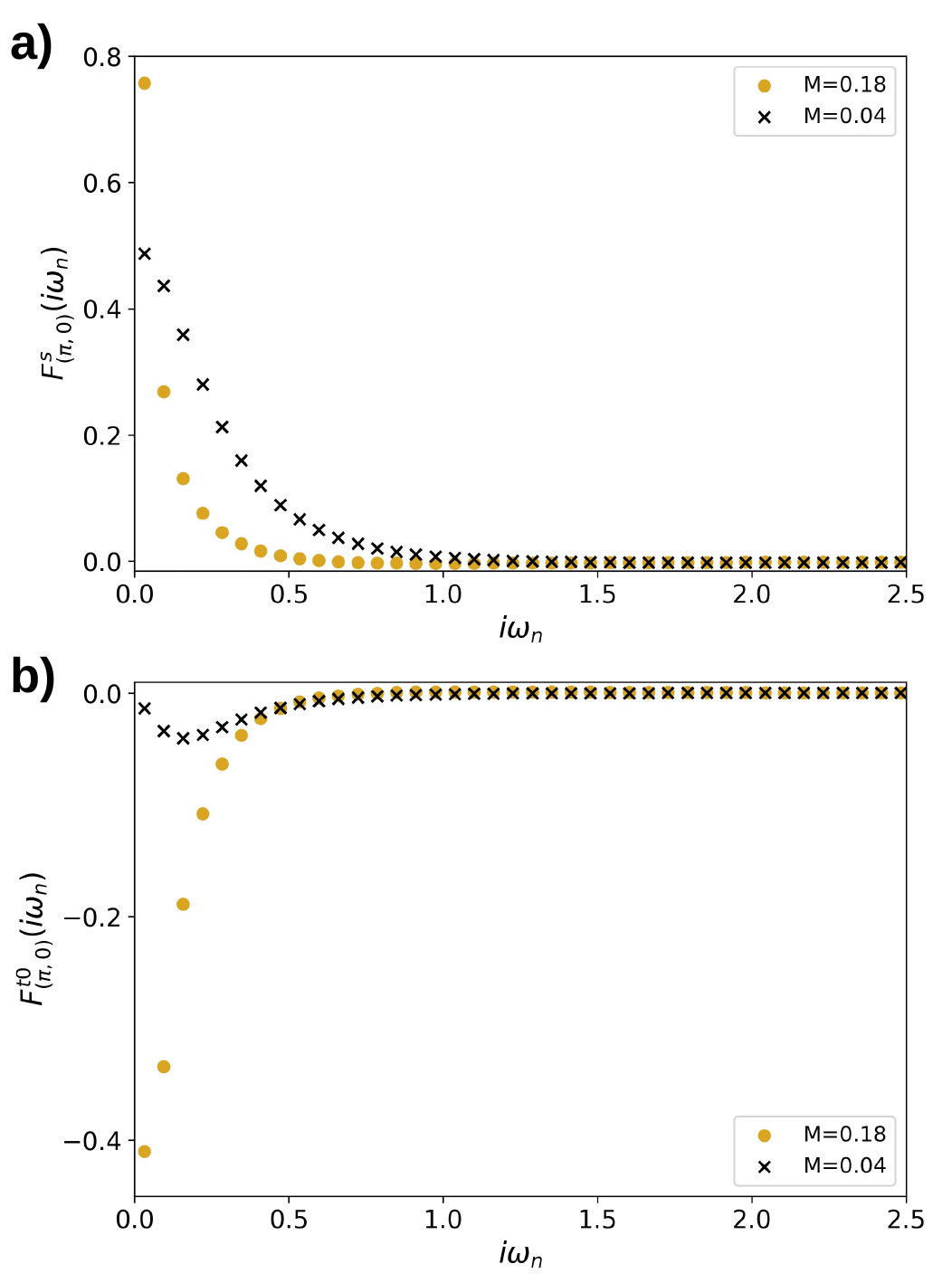}
    \caption{\textbf{a)} Singlet and \textbf{b)} $S_z=0$ triplet anomalous Green's function components in the positive Matsubara frequency grid at the antinode $\mathbf{k}=(\pi,0)$ for $M/t=0.04$ (black crosses) and $M/t=0.18$ (golden circles).}
    \label{fig:f_iwn}
\end{figure}

In Figure~\ref{fig:im_f_w} we complete the retarded-frequency characterization of the singlet and odd-frequency triplet components of $F^{\uparrow\downarrow}(k,\omega)$ by showing the imaginary parts at the antinodal momentum $\mathbf{k}=(\pi,0)$. They display the inverse symmetry with respect to the real parts, as expected. 

In Fig.~\ref{fig:f_iwn} we show the low-frequency behavior of the singlet and triplet components at the antinode $\mathbf{k}=(\pi,0)$ on the Matsubara axis (only positive frequencies are shown). The singlet component in~a) is purely real and the triplet component in~b) is purely imaginary. The singlet value at the first Matsubara point is enhanced for $M/t=0.04$ but decays more rapidly with $\omega_n$ if compared to $M/t=0.18$, while the magnitude of the triplet component grows strongly over the same low-frequency interval. This shows that the surviving anomalous correlations become increasingly concentrated at low energy as the exchange splitting increases inside the superconducting state. Both anomalous sectors vanish after the superconducting solution is lost, which happens in our model between $M/t=0.18$ and $M/t=0.19$.

\section{Quadratic BCS d-wave reference for the magnetic response}\label{bcs}

This Appendix highlights the part of the magnetic response that follows already from a quadratic Zeeman-split $d$-wave BCS-like Hamiltonian. It is important to recall that magnetic field-induced coexistence of even-frequency singlet and odd-frequency triplet correlations is a known consequence of broken time-reversal symmetry~\cite{matsumoto_koga_kusunose2012,linder_robinson2015,Perrin2020,Santos2020}. Likewise, paramagnetic contributions associated with odd-frequency pairing have been discussed in other settings~\cite{higashitani2013,higashitani2014}. Our goal is to establish a mean-field baseline against which the CDMFT results can be compared. We start from the same Nambu convention (Eq.~\ref{Nambu}) used to construct the Green's function formalism and its associated response functions. The calculations are performed a one-band model for both fixed chemical potential and fixed hole densities.

As a starting point, we consider a simplified uncorrelated model tight-binding dispersion $\xi_k$ in a square lattice dispersion associated to a $d$-wave gap $\Delta_k$ of the form
\begin{align}
    \xi_{\mathbf{k}}&=-2t(\cos k_x+\cos k_y) -4t'\cos k_x\cos k_y-\mu,\\
    \Delta_{\mathbf{k}}&=\frac{\Delta_0}{2}(\cos k_x-\cos k_y),
\end{align}
where hoppings were considered up to second nearest-neighbor ($t$ and $t'$), the $d$-wave amplitude is controlled by $\Delta_0$ and lattice occupation is defined by $\mu$. Here we utilize the same exchange-field convention as in the main text
\begin{equation}\label{eq:app_bcs_HM}
 H_{FM}=M\sum_{\mathbf{k}} \left(n_{\mathbf{k}\uparrow}-n_{\mathbf{k}\downarrow}\right),
\end{equation}
with $M$ the magnetic exchange field amplitude.

The Bogoliubov--de Gennes Hamiltonian and Green's function are
\begin{align}\label{eq:app_bcs_Ginv}
    H_{\mathrm{BdG}}(\mathbf{k}) &= M\tau_0+\xi_{\mathbf{k}}\tau_3+\Delta_{\mathbf{k}}\tau_1,\\
    G_{\mathrm{BdG}}^{-1}(\mathbf{k},i\omega_n)&=(i\omega_n-M)\tau_0-\xi_{\mathbf{k}}\tau_3 -\Delta_{\mathbf{k}}\tau_1,
\end{align}
with $\tau_i$ the corresponding two-dimensional Pauli matrices. Consequently, the first-order interaction vertexes associated to the magnetic response ($\Gamma_M$) and the density response ($\Gamma_n$) are
\begin{equation}\label{eq:app_bcs_vertices}
    \Gamma_M=\tau_0,\qquad \Gamma_n=\tau_3.
\end{equation}

Writing $E_{\mathbf{k}}^2=\xi_{\mathbf{k}}^2+\Delta_{\mathbf{k}}^2$, the anomalous propagator can be written as
\begin{equation}
    F(\mathbf{k},i\omega_n) = \frac{\Delta_{\mathbf{k}}}{(i\omega_n-M)^2-E_{\mathbf{k}}^2}.
\end{equation}
It is useful to derive explicitly its even and odd components (see Eq. \ref{eq:F_even_odd}) on the Matsubara axis:
\begin{align}\label{eq:app_bcs_even_odd}
    F_{\mathrm{even}}&=-\frac{\Delta_{\mathbf{k}}(\omega_n^2+E_{\mathbf{k}}^2-M^2)}{(\omega_n^2+E_{\mathbf{k}}^2-M^2)^2+(2M\omega_n)^2},\\
    F_{\mathrm{odd}}&=+\frac{2M\omega_n\Delta_{\mathbf{k}}}{(\omega_n^2+E_{\mathbf{k}}^2-M^2)^2+(2M\omega_n)^2}.
\end{align}
As discussed in the main text these components are real within a suitable gauge choice. This clearly 
shows that the $F_{\mathrm{even}}$ and $F_{\mathrm{odd}}$ are even and odd in $\omega_n$.

\subsection{Magnetic susceptibility $\chi$}
\begin{figure}
    \centering
    \includegraphics[width=0.9\linewidth]{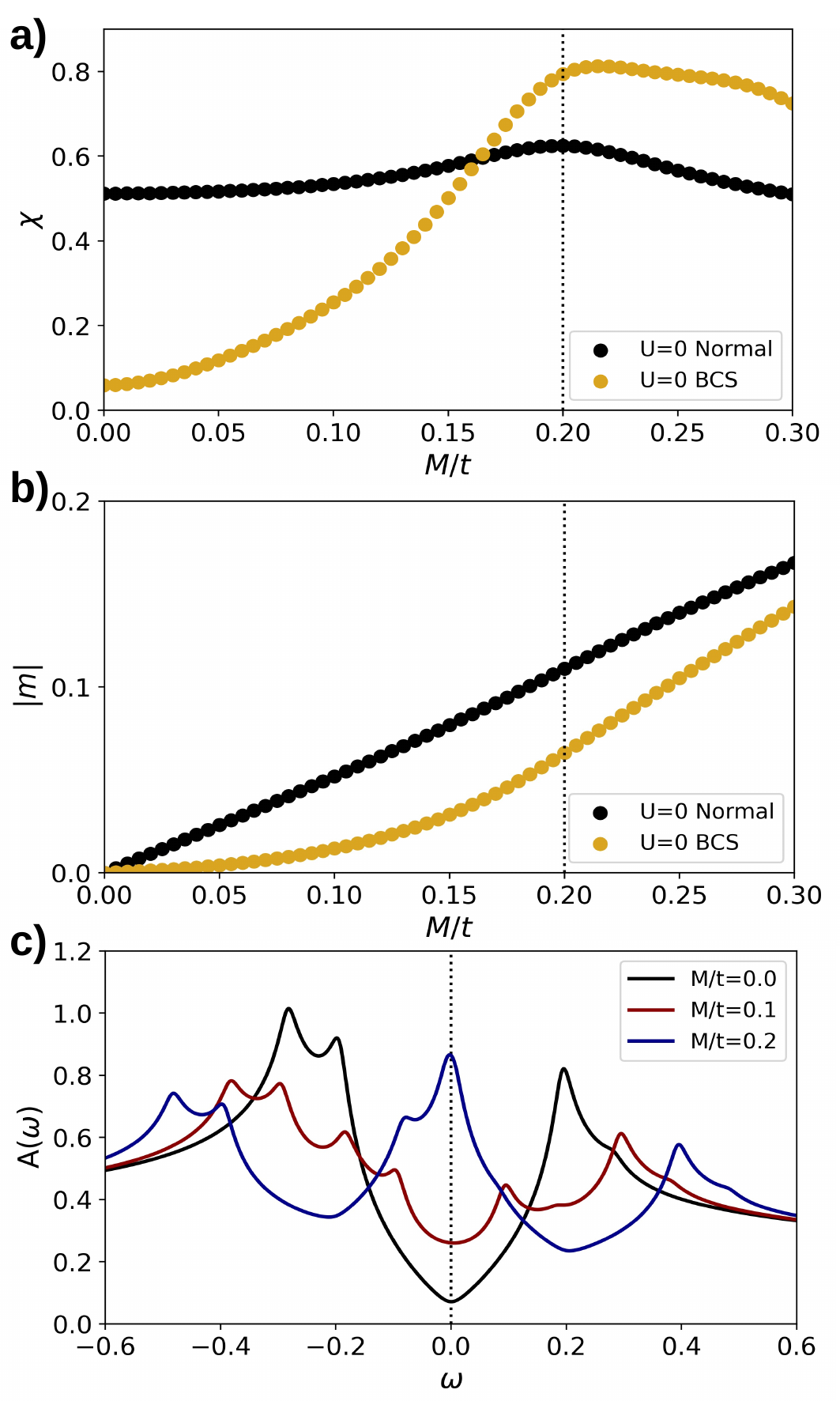}
    \caption{\textbf{a)} Magnetic susceptibility response and \textbf{b)} magnetization $M$ dependence in the $U=0$ BCS superconducting and normal state. \textbf{c)} Low energy spectral function for selected values of $M/t$. The total Nambu bubble is indistinguishable from $\partial|m|/\partial M$ evaluated directly. The dashed line indicates the $M=\Delta_0$ point. The continuation beyond the Pauli-limited stability range is included only as a reference.}
    \label{fig:bcs_norm}
\end{figure}

For $M/t>0$, the sign convention in Eq.~\eqref{eq:app_bcs_HM} gives $m=\langle n_\uparrow-n_\downarrow\rangle<0$. We therefore compare positive responses at a fixed chemical potential through
\begin{equation}
    \chi_\mu = \frac{\partial |m|}{\partial M}\bigg|_\mu = -\frac{\partial m}{\partial M}\bigg|_\mu.
\end{equation}
In this case, the static magnetic bare Kubo bubble is
\begin{equation}\label{eq:app_bcs_bubble}
    \chi_\mu^{\mathrm{bub}}=-\frac{T}{N_k}\sum_{\mathbf{k},\omega_n}\operatorname{Tr}\left[\tau_0G_{\mathrm{BdG}}(\mathbf{k},i\omega_n)\tau_0G_{\mathrm{BdG}}(\mathbf{k},i\omega_n)\right],
\end{equation}
which can be simplified as in Eq.~\eqref{eq:chi_decomp}
\begin{equation}\label{eq:app_bcs_decomp}
    \chi_\mu^{\mathrm{bub}} = \chi_G+\chi_{\mathrm{even}}+\chi_{\mathrm{odd}}.
\end{equation}

As the Hamiltonian is quadratic, there are no vertex corrections and the bubble is the complete response. An independent derivative can be then taken directly from
\begin{equation}
    m(M) = \frac{1}{N_k}\sum_{\mathbf{k}}\left[f(E_{\mathbf{k}}+M)-f(E_{\mathbf{k}}-M)\right],
\end{equation}
where $f(E)$ are the Fermionic distribution functions at a given energy $E$. The differentiation gives the direct susceptibility
\begin{equation}\label{eq:app_bcs_direct}
    \chi_\mu^{\mathrm{direct}} = \frac{1}{N_k}\sum_{\mathbf{k}}\left[-f'(E_{\mathbf{k}}+M)-f'(E_{\mathbf{k}}-M)\right].
\end{equation}
Equations~\eqref{eq:app_bcs_bubble} and \eqref{eq:app_bcs_direct} agree numerically to better than $4\times10^{-6}$ over the field interval shown in Fig. \ref{fig:bcs_norm} a).

The derivative in Eq.~\eqref{eq:app_bcs_direct} is taken at fixed chemical potential, whereas the CDMFT trajectory follows $\mu(M)$ to keep the hole density fixed at $p=0.07$. Fixing $p$ is equivalent to fixing the electron density $n=1-p$. We denote the fixed-density response by $\chi_p$ and define $\mathcal{M}=|m|$. The four static magnetic--density responses are
\begin{align}
    \chi_{MM}&=\left.\frac{\partial\mathcal{M}}{\partial M}\right|_\mu, &
    \chi_{Mn}&=\left.\frac{\partial\mathcal{M}}{\partial\mu}\right|_M,\\
    \chi_{nM}&=\left.\frac{\partial n}{\partial M}\right|_\mu, &
    \chi_{nn}&=\left.\frac{\partial n}{\partial\mu}\right|_M.
\end{align}
Along the fixed-density trajectory, the chemical potential obeys
\begin{align}\label{eq:app_density_constraint}
    \left.\frac{dn}{dM}\right|_p &= \chi_{nM}+\chi_{nn}\left.\frac{d\mu}{dM}\right|_p=0,\\
    \left.\frac{d\mu}{dM}\right|_p &= -\frac{\chi_{nM}}{\chi_{nn}}.
\end{align}
The total derivative of the magnetization at fixed density is, when including the chemical potential fluctuation correction, given by
\begin{equation}
    \chi_p = \left.\frac{d\mathcal{M}}{dM}\right|_p =\chi_{MM}-\frac{\chi_{Mn}\chi_{nM}}{\chi_{nn}}.
\end{equation}
With the sign convention of Eq.~\eqref{eq:app_bcs_HM}, the mixed derivatives are equal, $\chi_{Mn}=\chi_{nM}$. The subtracted term is therefore non-negative for a stable state with $\chi_{nn}>0$, and the fixed-density response cannot exceed the fixed-$\mu$ response evaluated at the same state.

Applying the same density constraint within the bubble approximation gives the response used in Sec.~\ref{subsec:beyond_bubble}:
\begin{equation}\label{eq:fixed_density_bubble_main}
    \chi_p^{\mathrm{bub}} = \chi_{MM}^{\mathrm{bub}} - \frac{\chi_{Mn}^{\mathrm{bub}}\chi_{nM}^{\mathrm{bub}}}{\chi_{nn}^{\mathrm{bub}}}.
\end{equation}
Here, $\chi_{MM}^{\mathrm{bub}}=\chi_\mu^{\mathrm{bub}}$, and the response-matrix entries are evaluated from the same one-particle Green's function:
\begin{align}\label{eq:chi_fix_def_bub}
    \chi_{MM}^{\mathrm{bub}} &= -\frac{T}{N_k}\sum_{\mathbf{k},\omega_n}\operatorname{Tr}\left[\tau_0\mathbf{G}\tau_0\mathbf{G}\right],\\
    \chi_{Mn}^{\mathrm{bub}}=\chi_{nM}^{\mathrm{bub}} &= +\frac{T}{N_k}\sum_{\mathbf{k},\omega_n}\operatorname{Tr}\left[\tau_0\mathbf{G}\tau_3\mathbf{G}\right],\\
    \chi_{nn}^{\mathrm{bub}} &= -\frac{T}{N_k}\sum_{\mathbf{k},\omega_n}\operatorname{Tr}\left[\tau_3\mathbf{G}\tau_3\mathbf{G}\right].
\end{align}
The arguments $(\mathbf{k},i\omega_n)$ of $\mathbf{G}$ are implicit. The positive sign of the mixed entry follows from using $\mathcal{M}=|m|=-m$ for $M>0$ with $H_{FM}=+M(n_\uparrow-n_\downarrow)$. For the quadratic reference, $\mathbf{G}=G_{\mathrm{BdG}}$ and, for the interacting bubble, $\mathbf{G}$ is the converged CDMFT Green's function. Interactions affect both these propagators and the trajectory $\mu(M)$ required to keep $p$ fixed. For the $U/t=8$ solutions, the relative correction to $\chi_{MM}^{\mathrm{bub}}$ is of order $10^{-3}$, comparable to the uncorrelated benchmarks of the quadratic BCS reference model. The difference between $\chi_{\mu}$ and $\chi_{p}$ is therefore too small for any practical comparison and it has been neglected in the main text. In particular this difference cannot explain the separation between the direct CDMFT susceptibility and the bubble one displayed in Fig.~\ref{fig:ord_evo_m} c).

\subsection{Susceptibility, magnetic moment and density of states}

In Figure~\ref{fig:bcs_norm} we present the $U=0$, $t=1$, $t'=-0.30$, $\mu=-1.00$, and $T=0.02$ comparison for the normal state ($\Delta_0=0.0$) and the BCS $d$-wave superconducting state ($\Delta_0=0.20$) with different values of $M$. In panel~a) we show the susceptibility $\chi$ evolution with $M$, indicating that the response in the superconducting state is smaller than the normal state for most of the values of $M$, before crossing the $M\approx\Delta$ line. This crossing before the critical $M$ is observed also in the bubble calculation from the CDMFT data in Fig.~\ref{fig:ord_evo_m} b), and ascribed to a peak appearing in the density of states. This observation is enforced by comparing the correlated CDMFT density of states of Fig.~\ref{fig:11} b) with the density of states of this non-interacting model in panel~c), at $M/t=0.2$ before the superconducting-to-normal phase transition. The integrated $|m|$ curves in panel~b) show though that the uncorrelated superconducting solution stays lower than the normal state for all the values of $M$ relevant for the superconducting correlated solution, in agreement with the bubble result for the correlated CDMFT case displayed on Fig.~\ref{fig:ord_evo_m} b). In panel~c) we highlight the local density of state of the superconducting solution for chosen values of $M$, demonstrating the destruction of the superconducting gap when moving from $M/t=0$ to $M/t=\Delta_0=0.2$ comparable with Fig.~\ref{fig:11}.

\section{\label{fs_evo} Detailed evolution of the MDC}
\begin{figure}
    \centering
    \includegraphics[width=1\linewidth]{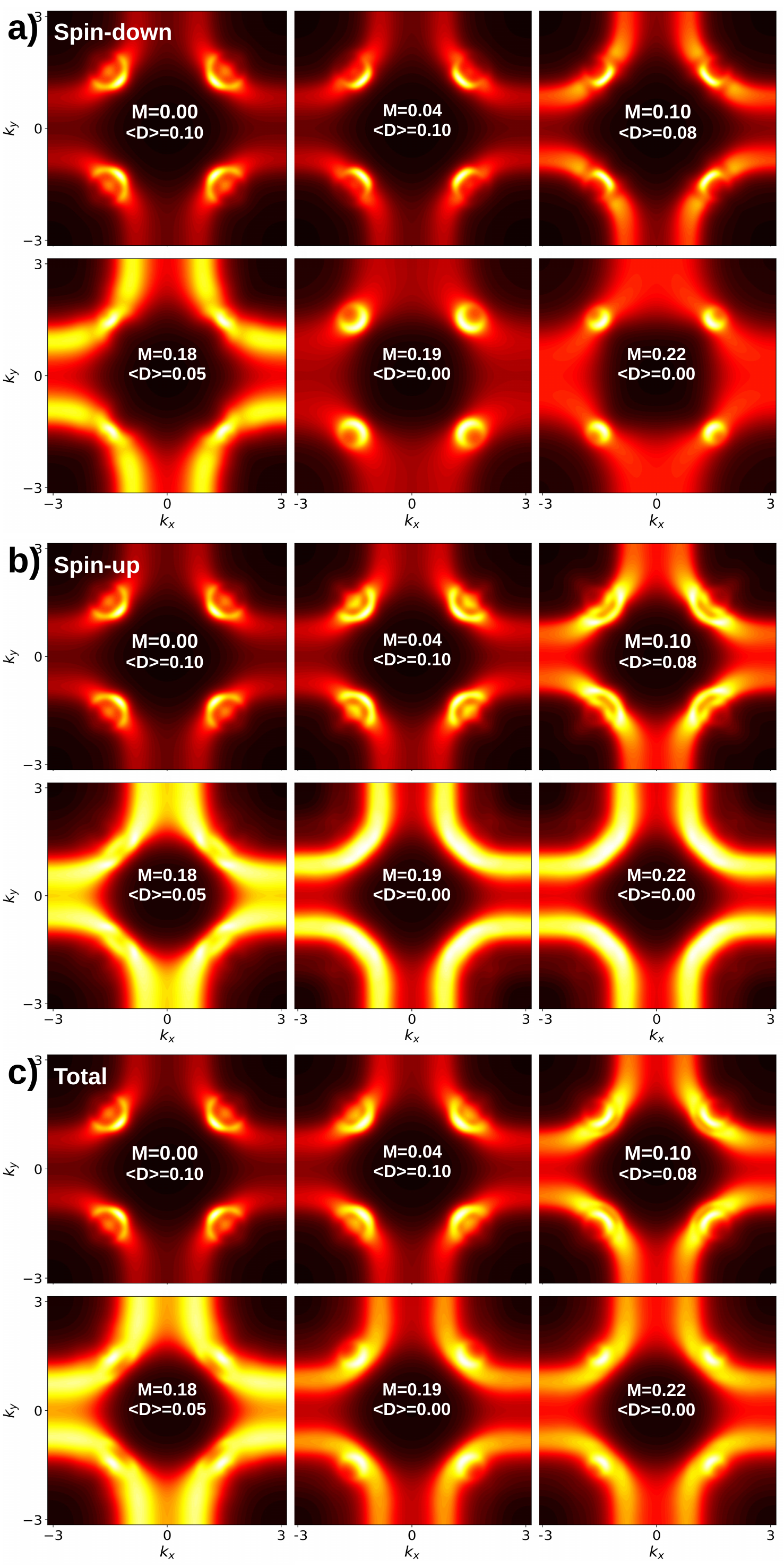}
    \caption{Momentum-resolved Fermi-level spectral weight at fixed hole doping $p=7\%$ for $M/t=0$, $0.04$, $0.10$, $0.18$, $0.19$, and $0.22$ (top row from left to right, followed by the bottom row from left to right, within each spin block). The value of $\langle D\rangle$ is indicated in every panel. \textbf{a)} Spin-down, \textbf{b)} spin-up, and \textbf{c)} total spectral weight.}
    \label{fig:M_evo_fs}
\end{figure} 

Figure~\ref{fig:M_evo_fs} follows the Fermi-level spectral dispersion $A_\sigma(\mathbf{k},0)$ as a function of the exchange field $M$ at fixed-doping. At $M/t=0$, the two spin projections are identical. Within the superconducting regime, increasing $M$ progressively differentiates the two spin sectors and redistributes weight along the underlying low-energy contours, while $\langle D\rangle$ decreases from approximately $0.10$ to $0.05$. The most pronounced reconstruction occurs between $M/t=0.18$ and $M/t=0.19$, where the superconducting solution disappears. In the spin-down channel (panel~a)), the extended low-energy structure present in the last superconducting solution is replaced by four pseudogap-like arcs confined to the nodal directions. In contrast, the spin-up channel (panel~b)) retains a much more extended and nearly continuous Fermi-liquid-like contour. The total spin spectral weight (panel~c)) therefore masks part of the pseudogap reconstruction, but it still displays the nodal--antinodal differentiation inherited from the strongly depleted spin-down sector. This abrupt spin selectivity cannot be then inferred from the total spectral weight alone. At $M/t=0.22$ in the normal phase, the anomalous weight is absent and the same qualitative differentiation persists, so the polarized pseudogap is not an isolated feature of the first normal-state point after the collapse of superconductivity.

\section{Polarized pseudogap phase}\label{pseudogap}
\begin{figure}[t!!]
    \centering
    \includegraphics[width=0.85\linewidth]{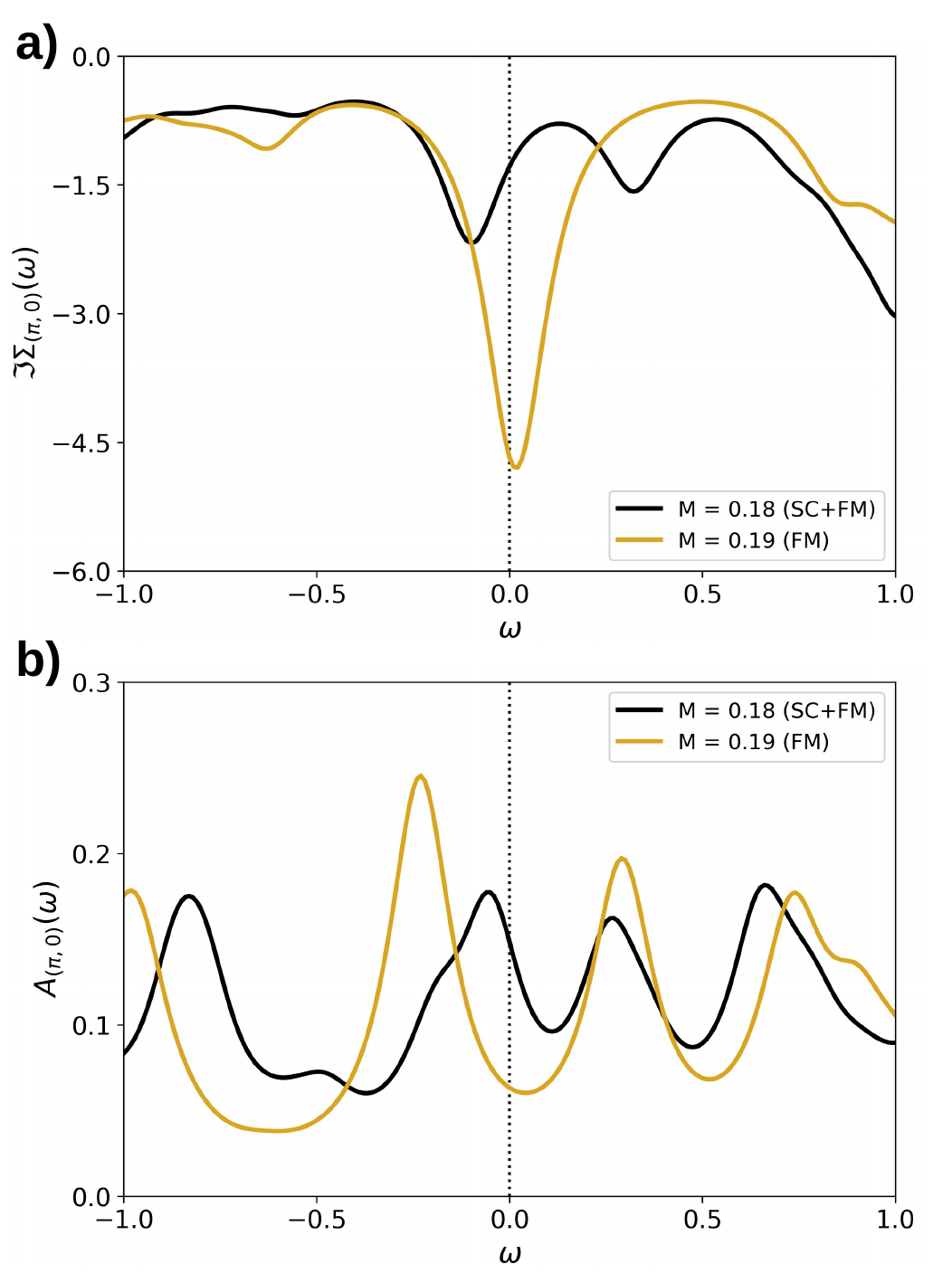}
    \caption{Spin-down antinodal \textbf{a)} imaginary part of the self-energy and \textbf{b)} spectral function at $\mathbf{k}=(\pi,0)$ close to the normal state transition. $M/t=0.18$ superconducting solution is indicated in black, and the $M/t=0.19$ first spin-polarized pseudogap solution is indicated in gold.}
    \label{fig:pg_int}
\end{figure}

For a more complete characterization of the pseudogap normal state, we can look in particular at the self-energy and the corresponding spectral function at low-frequency for the spin specie that shows the pseudogap character. In Fig.~\ref{fig:pg_int} a) we compare the spin-down sector self-energy and spectral function at $\mathbf{k}=(\pi,0)$ immediately before and after the collapse of superconductivity. At $M/t=0.18$, the system is still superconducting but on the verge of the transition. Close to $\omega=0$ the self-energy shows to be close to the Fermi-liquid-like $~\omega^2$ minimum, displaying a small peak. This anomalous shape corresponds to the maximum in the density of states (panel~b)), rising from the spin-spitted gap discussed in Fig.~\ref{fig:11} b). At $M/t=0.19$, however, $\langle D\rangle=0$, and the negative imaginary self-energy develops a strong pole-like structure close to $\omega=0$, corresponding to the characteristic depression in the density of states (panel~b)) typical of the strongly correlated normal-state pseudogap found in previous CDMFT studies~\cite{sakai_s_wave_pseudogap_2013}. This inequitably confirms the pseudogap nature of the spectral features that we found in the spin down sector in the $M$-induced normal state.


\bibliography{ref}

@article{Santos2020,
  title = {Odd-frequency superconductivity in dilute magnetic superconductors},
  author = {Santos, Fl{\'a}vio L. N. and Perrin, Vivien and Jamet, Fran{\c{c}}ois and Civelli, Marcello and Simon, Pascal and Aguiar, Maria C. O. and Miranda, Eduardo and Rozenberg, Marcelo J.},
  journal = {Phys. Rev. Research},
  volume = {2},
  number = {3},
  pages = {033229},
  year = {2020},
  month = {Aug},
  publisher = {American Physical Society},
  doi = {10.1103/PhysRevResearch.2.033229}
}

@article{Perrin2020,
  title = {Unveiling Odd-Frequency Pairing around a Magnetic Impurity in a Superconductor},
  author = {Perrin, Vivien and Santos, Fl{\'a}vio L. N. and M{\'e}nard, Gerbold C. and Brun, Christophe and Cren, Tristan and Civelli, Marcello and Simon, Pascal},
  journal = {Phys. Rev. Lett.},
  volume = {125},
  number = {11},
  pages = {117003},
  year = {2020},
  month = {Sep},
  publisher = {American Physical Society},
  doi = {10.1103/PhysRevLett.125.117003}
}

@article{gas_2d,
title = {Metallic behavior in {STO}/{LAO} heterostructures with non-uniformly atomic interfaces},
journal = {Materials Today Communications},
volume = {24},
pages = {101339},
year = {2020},
issn = {2352-4928},
doi = {https://doi.org/10.1016/j.mtcomm.2020.101339},
url = {https://www.sciencedirect.com/science/article/pii/S2352492820323503},
author = {Rafael A.C. Amoresi and Leonélio Cichetto and Amanda F. Gouveia and Yormary N. Colmenares and Marcio D. Teodoro and Gilmar E. Marques and Elson Longo and Alexandre Z. Simões and Juan Andrés and Adenilson J. Chiquito and Maria A. Zaghete},
language = {english}
}

@article{sano2026,
  author  = {Sano, Tomoya and Tabata, Kota and Sasaki, Akihiro and Asano, Yasuhiro},
  title   = {Low-temperature anomaly and anisotropy of critical magnetic fields in transition metal dichalcogenide superconductors},
  journal = {Physical Review B},
  volume  = {114},
  pages   = {024506},
  year    = {2026},
  doi     = {10.1103/y9k6-221l}
}

@article{sakai_s_wave_pseudogap_2013,
  author  = {Sakai, Shiro and Blanc, S. and Civelli, Marcello
             and Gallais, Y. and Cazayous, M. and M{\'e}asson, M.-A.
             and Wen, J. S. and Xu, Z. J. and Gu, G. D.
             and Sangiovanni, G. and Motome, Y. and Held, K.
             and Sacuto, A. and Georges, A. and Imada, M.},
  title   = {Raman-Scattering Measurements and Theory of the
             Energy-Momentum Spectrum for Underdoped
             {Bi$_2$Sr$_2$CaCuO$_{8+\delta}$} Superconductors:
             Evidence of an {$s$}-Wave Structure for the Pseudogap},
  journal = {Physical Review Letters},
  volume  = {111},
  number  = {10},
  pages   = {107001},
  year    = {2013},
  doi     = {10.1103/PhysRevLett.111.107001}
}

@article{CHOU2024158739,
title = {Controllable spin-triplet superconductivity states and enhanced non-dissipation spin-polarized supercurrents in YBa2Cu3O7/La0.67Sr0.33MnO3 interfaces},
journal = {Applied Surface Science},
volume = {644},
pages = {158739},
year = {2024},
issn = {0169-4332},
doi = {https://doi.org/10.1016/j.apsusc.2023.158739},
url = {https://www.sciencedirect.com/science/article/pii/S0169433223024194},
author = {Hsiung Chou and S.J. Sun and Kung-Shang Yang and G.D. Dwivedi and Chiu-Hao Chen and S.L. Cheng and J.G. Lin and J.W. Chiou and Y.Y. Chin and H.J. Lin and V.I. Grebennikov}
}

@article{Liechtenstein_1995,
title = {Density-functional theory and strong interactions: {O}rbital ordering in {M}ott-{H}ubbard insulators},
author = {Liechtenstein, A. I. and Anisimov, V. I. and Zaanen, J.},
journal = {Phys. Rev. B},
volume = {52},
issue = {8},
pages = {R5467--R5470},
year = {1995},
month = {Aug},
publisher = {American Physical Society},
doi = {10.1103/PhysRevB.52.R5467},
url = {https://link.aps.org/doi/10.1103/PhysRevB.52.R5467}
}

@article{Damascelli_2008,
  title = {Theory of Fermi-surface pockets and correlation effects in underdoped {Y}{B}a$_2${C}u$_{3}${O}$_{6.5}$},
  author = {Elfimov, I. S. and Sawatzky, G. A. and Damascelli, A.},
  journal = {Phys. Rev. B},
  volume = {77},
  issue = {6},
  pages = {060504},
  numpages = {4},
  year = {2008},
  month = {Feb},
  publisher = {American Physical Society},
  doi = {10.1103/PhysRevB.77.060504},
  url = {https://link.aps.org/doi/10.1103/PhysRevB.77.060504}
}

@article{CHEN2004295,
title = {Suppression of superconductivity in YBCO/LCMO superlattices},
journal = {Solid State Communications},
volume = {131},
number = {5},
pages = {295-299},
year = {2004},
issn = {0038-1098},
doi = {https://doi.org/10.1016/j.ssc.2004.05.028},
url = {https://www.sciencedirect.com/science/article/pii/S0038109804004375},
author = {F Chen and B Gorshunov and G Cristiani and H.-U Habermeier and M Dressel}
}

@article{wien2wannier,
title = {Wien2wannier: From linearized augmented plane waves to maximally localized Wannier functions},
journal = {Computer Physics Communications},
volume = {181},
number = {11},
pages = {1888-1895},
year = {2010},
issn = {0010-4655},
doi = {https://doi.org/10.1016/j.cpc.2010.08.005},
url = {https://www.sciencedirect.com/science/article/pii/S0010465510002948},
author = {Jan Kuneš and Ryotaro Arita and Philipp Wissgott and Alessandro Toschi and Hiroaki Ikeda and Karsten Held}
}

@article{kancharla_sc_af_2008,
  author  = {Kancharla, S. S. and Kyung, B. and S{\'e}n{\'e}chal, D.
             and Civelli, M. and Capone, M. and Kotliar, G.
             and Tremblay, A.-M. S.},
  title   = {Anomalous superconductivity and its competition
             with antiferromagnetism in doped Mott insulators},
  journal = {Physical Review B},
  volume  = {77},
  number  = {18},
  pages   = {184516},
  year    = {2008},
  doi     = {10.1103/PhysRevB.77.184516}
}

@article{senechal_bath_optimization_2010,
  author  = {S{\'e}n{\'e}chal, David},
  title   = {Bath optimization in the Cellular Dynamical Mean Field Theory},
  journal = {Physical Review B},
  volume  = {81},
  pages   = {235125},
  year    = {2010},
  doi     = {10.1103/PhysRevB.81.235125}
}

@article{kotliar_cdmft_prl,
  author  = {Kotliar, Gabriel and Savrasov, Sergey Y. and P{\'a}lsson, Gunnar and Biroli, Giulio},
  title   = {Cellular dynamical mean field approach to strongly correlated systems},
  journal = {Physical Review Letters},
  volume  = {87},
  pages   = {186401},
  year    = {2001},
  doi     = {10.1103/PhysRevLett.87.186401}
}

@article{maier_cluster_rmp,
  author  = {Maier, Thomas and Jarrell, Mark and Pruschke, Thomas and Hettler, Matthias H.},
  title   = {Quantum cluster theories},
  journal = {Reviews of Modern Physics},
  volume  = {77},
  pages   = {1027--1080},
  year    = {2005},
  doi     = {10.1103/RevModPhys.77.1027}
}

@article{kotliar_dmft_rmp_2006,
  author  = {Kotliar, Gabriel and Savrasov, Sergey Y. and Haule, Kristjan and Oudovenko, Victor S. and Parcollet, Olivier and Marianetti, Chris A.},
  title   = {Electronic structure calculations with dynamical mean-field theory},
  journal = {Reviews of Modern Physics},
  volume  = {78},
  pages   = {865--951},
  year    = {2006},
  doi     = {10.1103/RevModPhys.78.865}
}

@article{tremblay_lowtemp_2006,
  author  = {Tremblay, A.-M. S. and Kyung, B. and S{\'e}n{\'e}chal, D.},
  title   = {Pseudogap and high-temperature superconductivity from weak to strong coupling. Towards a quantitative theory},
  journal = {Low Temperature Physics},
  volume  = {32},
  pages   = {424--451},
  year    = {2006},
  doi     = {10.1063/1.2199446}
}

@article{lee_nagaosa_wen_2006,
  author    = {Lee, Patrick A. and Nagaosa, Naoto and Wen, Xiao-Gang},
  title     = {Doping a Mott insulator: Physics of high-temperature superconductivity},
  journal   = {Reviews of Modern Physics},
  volume    = {78},
  pages     = {17--85},
  year      = {2006},
  doi       = {10.1103/RevModPhys.78.17},
  url       = {https://doi.org/10.1103/RevModPhys.78.17}
}

@article{qin_hubbard_2022,
  author    = {Qin, Mingpu and Sch{\"a}fer, Thomas and Andergassen, Sabine and Corboz, Philippe and Gull, Emanuel},
  title     = {The Hubbard Model: A Computational Perspective},
  journal   = {Annual Review of Condensed Matter Physics},
  volume    = {13},
  pages     = {275--302},
  year      = {2022},
  doi       = {10.1146/annurev-conmatphys-090921-033948},
  url       = {https://doi.org/10.1146/annurev-conmatphys-090921-033948}
}

@article{zhou_natphys_2025,
  author  = {Zhou, R. and Vinograd, I. and Hirata, M. and Wu, T. and Mayaffre, H. and Kr{\"a}mer, S. and Hardy, W. N. and Liang, R. and Bonn, D. A. and Loew, T. and Porras, J. and Keimer, B. and Julien, M.-H.},
  title   = {Signatures of two gaps in the spin susceptibility of a cuprate superconductor},
  journal = {Nature Physics},
  volume  = {21},
  pages   = {97--103},
  year    = {2025},
  doi     = {10.1038/s41567-024-02692-w}
}

@article{fradkin_intertwined_orders_2015,
  author  = {Fradkin, Eduardo and Kivelson, Steven A. and Tranquada, John M.},
  title   = {Colloquium: Theory of intertwined orders in high temperature superconductors},
  journal = {Reviews of Modern Physics},
  volume  = {87},
  pages   = {457--482},
  year    = {2015},
  doi     = {10.1103/RevModPhys.87.457}
}

@article{comin_damascelli_2016,
  author  = {Comin, Riccardo and Damascelli, Andrea},
  title   = {Resonant X-Ray Scattering Studies of Charge Order in Cuprates},
  journal = {Annual Review of Condensed Matter Physics},
  volume  = {7},
  pages   = {369--405},
  year    = {2016},
  doi     = {10.1146/annurev-conmatphys-031115-011401}
}

@article{hayden_tranquada_2024,
  author  = {Hayden, Stephen M. and Tranquada, John M.},
  title   = {Charge Correlations in Cuprate Superconductors},
  journal = {Annual Review of Condensed Matter Physics},
  volume  = {15},
  pages   = {215--238},
  year    = {2024},
  doi     = {10.1146/annurev-conmatphys-032922-094430}
}

@article{bourges_loop_currents_2021,
  author  = {Bourges, Philippe and Sidis, Yvan},
  title   = {Loop currents in quantum matter},
  journal = {Comptes Rendus Physique},
  volume  = {22},
  pages   = {59--89},
  year    = {2021},
  doi     = {10.5802/crphys.84}
}

@article{berezinskii1974,
  author  = {Berezinskii, V. L.},
  title   = {New model of the anisotropic phase of superfluid $^3$He},
  journal = {JETP Letters},
  volume  = {20},
  number  = {9},
  pages   = {287--289},
  year    = {1974}
}

@article{linder_balatsky2019,
  author  = {Linder, Jacob and Balatsky, Alexander V.},
  title   = {Odd-frequency superconductivity},
  journal = {Reviews of Modern Physics},
  volume  = {91},
  pages   = {045005},
  year    = {2019},
  doi     = {10.1103/RevModPhys.91.045005}
}

@article{bergeret_volkov_efetov2005,
  author  = {Bergeret, F. S. and Volkov, A. F. and Efetov, K. B.},
  title   = {Odd triplet superconductivity and related phenomena in superconductor-ferromagnet structures},
  journal = {Reviews of Modern Physics},
  volume  = {77},
  pages   = {1321--1373},
  year    = {2005},
  doi     = {10.1103/RevModPhys.77.1321}
}

@article{buzdin2005,
  author  = {Buzdin, A. I.},
  title   = {Proximity effects in superconductor-ferromagnet heterostructures},
  journal = {Reviews of Modern Physics},
  volume  = {77},
  pages   = {935},
  year    = {2005},
  doi     = {10.1103/RevModPhys.77.935}
}

@article{qcm,
	title={{Pyqcm: An open-source Python library for quantum cluster methods}},
	author={Théo N. Dionne and Alexandre Foley and Moïse Rousseau and David Sénéchal},
	journal={SciPost Phys. Codebases},
	pages={23},
	year={2023},
	publisher={SciPost},
	doi={10.21468/SciPostPhysCodeb.23},
	url={https://scipost.org/10.21468/SciPostPhysCodeb.23},
}

@article{meu_int_dft,
  title = {Interface and thickness effects in {L}a$_{2/3}${S}r$_{1/3}${M}n{O}$_{3}$/{Y}{B}a$_{2}${C}u$_{3}${O}$_{7}$ superlattices},
  author = {Lima, V. A. M. and Aguiar, M. C. O. and Plumb, N. C. and Radovic, M. and Brito, W. H.},
  journal = {Phys. Rev. B},
  volume = {109},
  issue = {4},
  pages = {045128},
  numpages = {12},
  year = {2024},
  month = {Jan},
  publisher = {American Physical Society},
  doi = {10.1103/PhysRevB.109.045128},
  url = {https://link.aps.org/doi/10.1103/PhysRevB.109.045128}
}

@Article{Soltan2023,
author={Soltan, S. and Macke, S. and Ilse, S. E. and Pennycook, T. and Zhang, Z. L. and Christiani, G. and Benckiser, E. and Sch{\"u}tz, G. and Goering, E.},
title={Ferromagnetic order controlled by the magnetic interface of LaNiO3/La2/3Ca1/3MnO3 superlattices},
journal={Scientific Reports},
year={2023},
month={Mar},
day={08},
volume={13},
number={1},
pages={3847},
issn={2045-2322},
doi={10.1038/s41598-023-30814-6},
url={https://doi.org/10.1038/s41598-023-30814-6}
}

@article{tb_weber,
doi = {10.1209/0295-5075/100/37001},
url = {https://dx.doi.org/10.1209/0295-5075/100/37001},
year = {2012},
month = {nov},
publisher = {EDP Sciences, IOP Publishing and Società Italiana di Fisica},
volume = {100},
number = {3},
pages = {37001},
author = {Weber, C. and Yee, C. and Haule, K. and Kotliar, G.},
title = {Scaling of the transition temperature of hole-doped cuprate superconductors with the charge-transfer energy},
journal = {Europhysics Letters}
}

@article{tb_sacha,
author = {O.K. Andersen and A.I. Liechtenstein and O. Jepsen and F. Paulsen},
title = {LDA energy bands, low-energy hamiltonians, t', t", t$_{\bot}$(k), and J$\bot$},
journal = {Journal of Physics and Chemistry of Solids},
volume = {56},
number = {12},
pages = {1573-1591},
year = {1995},
note = {Proceedings of the Conference on Spectroscopies in Novel Superconductors},
issn = {0022-3697},
doi = {https://doi.org/10.1016/0022-3697(95)00269-3},
url = {https://www.sciencedirect.com/science/article/pii/0022369795002693}
}

@article{xas_chg_transf,
author = {J. Chakhalian  and J. W. Freeland  and H.-U. Habermeier  and G. Cristiani  and G. Khaliullin  and M. van Veenendaal  and B. Keimer },
title = {Orbital {R}econstruction and {C}ovalent {B}onding at an {O}xide {I}nterface},
journal = {Science},
volume = {318},
number = {5853},
pages = {1114-1117},
year = {2007},
doi = {10.1126/science.1149338},
URL = {https://www.science.org/doi/abs/10.1126/science.1149338},
eprint = {https://www.science.org/doi/pdf/10.1126/science.1149338}
}

@Article{dead_layer,
author={Chakhalian, J. and Freeland, J. W. and Srajer, G. and Strempfer, J. and Khaliullin, G. and Cezar, J. C. and Charlton, T. and Dalgliesh, R. and Bernhard, C. and Cristiani, G. and Habermeier, H.-U. and  Keimer, B.},
title={Magnetism at the interface between ferromagnetic and superconducting oxides},
journal={Nature Physics},
year={2006},
day={01},
volume={2},
number={4},
pages={244-248},
issn={1745-2481},
doi={10.1038/nphys272},
url={https://doi.org/10.1038/nphys272}
}

@article{sic_ybco_lsmo,
author={Gray, B. A. and Middey, S. and Conti, G. and Gray, A. X. and Kuo, C.-T. and Kaiser, A. M. and Ueda, S. and  Kobayashi, K. and Meyers, D. and Kareev, M. and Tung, I. C. and Liu, Jian and Fadley, C. S. and Chakhalian, J. and Freeland, J. W.},
title = {Superconductor to {M}ott insulator transition in {Y}{B}a$_2${C}u$_3${O}$_7$/{L}a{C}a{M}n{O}$_3$ heterostructures},
journal={Scientific Reports},
year={2016},
month={Sep},
day={15},
volume={6},
number={1},
pages={33184},
issn={2045-2322},
doi={10.1038/srep33184},
url={https://doi.org/10.1038/srep33184}
}

@article{wien2k,
    author = {Blaha, Peter and Schwarz, Karlheinz and Tran, Fabien and Laskowski, Robert and Madsen, Georg K. H. and Marks, Laurence D.},
    title = {WIEN2k: An APW+lo program for calculating the properties of solids},
    journal = {The Journal of Chemical Physics},
    volume = {152},
    number = {7},
    pages = {074101},
    year = {2020},
    month = {02},
    issn = {0021-9606},
    doi = {10.1063/1.5143061},
    url = {https://doi.org/10.1063/1.5143061},
}

@article{wannier90,
	doi = {10.1088/1361-648x/ab51ff},
	url = {https://doi.org/10.1088%2F1361-648x%2Fab51ff},
	year = 2020,
	month = {jan},
	publisher = {{IOP} Publishing},
	volume = {32},
	number = {16},
	pages = {165902},
	author = {Giovanni Pizzi and Valerio Vitale and Ryotaro Arita and Stefan Blügel and Frank Freimuth and Guillaume G{\'{e}}ranton and Marco Gibertini and Dominik Gresch and Charles Johnson and Takashi Koretsune and Julen Iba{\~{n}}ez-Azpiroz and Hyungjun Lee and Jae-Mo Lihm and Daniel Marchand and Antimo Marrazzo and Yuriy Mokrousov and Jamal I Mustafa and Yoshiro Nohara and Yusuke Nomura and Lorenzo Paulatto and Samuel Ponc{\'{e}} and Thomas Ponweiser and Junfeng Qiao and Florian Thöle and Stepan S Tsirkin and Ma{\l}gorzata Wierzbowska and Nicola Marzari and David Vanderbilt and Ivo Souza and Arash A Mostofi and Jonathan R Yates},
	title = {Wannier90 as a community code: new features and applications},
	journal = {Journal of Physics: Condensed Matter}
}

@article{primme,
author = {Stathopoulos, Andreas and McCombs, James R.},
title = {PRIMME: preconditioned iterative multimethod eigensolver—methods and software description},
year = {2010},
issue_date = {April 2010},
publisher = {Association for Computing Machinery},
address = {New York, NY, USA},
volume = {37},
number = {2},
issn = {0098-3500},
url = {https://doi.org/10.1145/1731022.1731031},
doi = {10.1145/1731022.1731031},
journal = {ACM Trans. Math. Softw.},
month = apr,
articleno = {21},
numpages = {30}
}

@article{senechal_diag_afm_sc,
  title = {Coexistence of superconductivity and antiferromagnetism in the Hubbard model for cuprates},
  author = {Foley, A. and Verret, S. and Tremblay, A.-M. S. and S\'en\'echal, D.},
  journal = {Phys. Rev. B},
  volume = {99},
  issue = {18},
  pages = {184510},
  numpages = {11},
  year = {2019},
  month = {May},
  publisher = {American Physical Society},
  doi = {10.1103/PhysRevB.99.184510},
  url = {https://link.aps.org/doi/10.1103/PhysRevB.99.184510}
}

@article{matsumoto_koga_kusunose2012,
  author  = {Matsumoto, Masashige and Koga, Mikito and Kusunose, Hiroaki},
  title   = {Coexistence of Even- and Odd-Frequency Superconductivities Under Broken Time-Reversal Symmetry},
  journal = {Journal of the Physical Society of Japan},
  volume  = {81},
  pages   = {033702},
  year    = {2012},
  doi     = {10.1143/JPSJ.81.033702},
  language = {english}
}

@article{tri_cup_man_1,
  title = {X-ray absorption spectroscopy study of the electronic and magnetic proximity effects in ${\mathrm{YBa}}_{2}{\mathrm{Cu}}_{3}{\mathrm{O}}_{7}/{\mathrm{La}}_{2/3}{\mathrm{Ca}}_{1/3}{\mathrm{MnO}}_{3}$ and ${\mathrm{La}}_{2\ensuremath{-}x}{\mathrm{Sr}}_{x}{\mathrm{CuO}}_{4}/{\mathrm{La}}_{2/3}{\mathrm{Ca}}_{1/3}{\mathrm{MnO}}_{3}$ multilayers},
  author = {Uribe-Laverde, M. A. and Das, S. and Sen, K. and Marozau, I. and Perret, E. and Alberca, A. and Heidler, J. and Piamonteze, C. and Merz, M. and Nagel, P. and Schuppler, S. and Munzar, D. and Bernhard, C.},
  journal = {Phys. Rev. B},
  volume = {90},
  issue = {20},
  pages = {205135},
  numpages = {7},
  year = {2014},
  month = {Nov},
  publisher = {American Physical Society},
  doi = {10.1103/PhysRevB.90.205135},
  url = {https://link.aps.org/doi/10.1103/PhysRevB.90.205135}
}

@article{tri_cup_man_3,
    author = {Sarkar, Subhrangsu and Capu, Roxana and Pashkevich, Yurii G and Knobel, Jonas and Cantarino, Marli R and Nag, Abhishek and Kummer, Kurt and Betto, Davide and Sant, Roberto and Nicholson, Christopher W and Khmaladze, Jarji and Zhou, Ke-Jin and Brookes, Nicholas B and Monney, Claude and Bernhard, Christian},
    title = {Composite antiferromagnetic and orbital order with altermagnetic properties at a cuprate/manganite interface},
    journal = {PNAS Nexus},
    volume = {3},
    number = {4},
    pages = {pgae100},
    year = {2024},
    month = {03},
    issn = {2752-6542},
    doi = {10.1093/pnasnexus/pgae100},
    url = {https://doi.org/10.1093/pnasnexus/pgae100},
    eprint = {https://academic.oup.com/pnasnexus/article-pdf/3/4/pgae100/57469478/pgae100.pdf},
}

@article{Sanchez-Manzano2022,
title = "Extremely long-range, high-temperature Josephson coupling across a half-metallic ferromagnet",
author = "Sanchez-Manzano, D and Mesoraca, S and Cuellar, F A and Cabero, M and Rouco, V and Orfila, G and Palermo, X and Balan, A and Marcano, L and Sander, A and Rocci, M and Garcia-Barriocanal, J and Gallego, F and Tornos, J and Rivera, A and Mompean, F and Garcia-Hernandez, M and Gonzalez-Calbet, J M and Leon, C and Valencia, S and Feuillet-Palma, C and Bergeal, N and Buzdin, A I and Lesueur, J and Villegas, Javier E and Santamaria, J",
journal  = "Nature Materials",
volume   =  21,
number   =  2,
pages    = "188--194",
month    =  feb,
year     =  2022
}

@article{marcello_pg_sc,
  title = {Direct connection between Mott insulators and $d$-wave high-temperature superconductors revealed by continuous evolution of self-energy poles},
  author = {Sakai, Shiro and Civelli, Marcello and Imada, Masatoshi},
  journal = {Phys. Rev. B},
  volume = {98},
  issue = {19},
  pages = {195109},
  numpages = {14},
  year = {2018},
  month = {Nov},
  publisher = {American Physical Society},
  doi = {10.1103/PhysRevB.98.195109},
  url = {https://link.aps.org/doi/10.1103/PhysRevB.98.195109}
}

@article{linder_robinson2015,
  author  = {Linder, Jacob and Robinson, Jason W. A.},
  title   = {Strong odd-frequency correlations in fully gapped Zeeman-split superconductors},
  journal = {Scientific Reports},
  volume  = {5},
  pages   = {15483},
  year    = {2015},
  doi     = {10.1038/srep15483},
  language = {english}
}

@article{ybco_M,
doi = {10.1088/0953-2048/29/7/075002},
url = {https://dx.doi.org/10.1088/0953-2048/29/7/075002},
year = {2016},
month = {may},
publisher = {IOP Publishing},
volume = {29},
number = {7},
pages = {075002},
author = {Golovchanskiy, Igor A and Pan, Alexey V and George, Jonathan and Wells, Frederick S and Fedoseev, Sergey A and Rozenfeld, Anatoly},
title = {Vibration effect on magnetization and critical current density of superconductors},
journal = {Superconductor Science and Technology}
}

@article{ybco_dx2y2,
  title = {Superconducting electronic state in optimally doped ${\text{YBa}}_{2}{\text{Cu}}_{3}{\text{O}}_{7\ensuremath{-}\ensuremath{\delta}}$ observed with laser-excited angle-resolved photoemission spectroscopy},
  author = {Okawa, M. and Ishizaka, K. and Uchiyama, H. and Tadatomo, H. and Masui, T. and Tajima, S. and Wang, X.-Y. and Chen, C.-T. and Watanabe, S. and Chainani, A. and Saitoh, T. and Shin, S.},
  journal = {Phys. Rev. B},
  volume = {79},
  issue = {14},
  pages = {144528},
  numpages = {9},
  year = {2009},
  month = {Apr},
  publisher = {American Physical Society},
  doi = {10.1103/PhysRevB.79.144528},
  url = {https://link.aps.org/doi/10.1103/PhysRevB.79.144528}
}

@article{vasp1,
title = {Ab initio molecular dynamics for liquid metals},
author = {Kresse, G. and Hafner, J.},
journal = {Phys. Rev. B},
volume = {47},
issue = {1},
pages = {558--561},
numpages = {0},
year = {1993},
month = {Jan},
publisher = {American Physical Society},
doi = {10.1103/PhysRevB.47.558},
url = {https://link.aps.org/doi/10.1103/PhysRevB.47.558}
}

@article{vasp2,
title = {Efficiency of ab-initio total energy calculations for metals and semiconductors using a plane-wave basis set},
journal = {Computational Materials Science},
volume = {6},
number = {1},
pages = {15-50},
year = {1996},
issn = {0927-0256},
doi = {https://doi.org/10.1016/0927-0256(96)00008-0},
url = {https://www.sciencedirect.com/science/article/pii/0927025696000080},
author = {G. Kresse and J. Furthmüller}
}

@article{vasp3,
title = {Efficient iterative schemes for ab initio total-energy calculations using a plane-wave basis set},
author = {Kresse, G. and Furthm\"uller, J.},
journal = {Phys. Rev. B},
volume = {54},
issue = {16},
pages = {11169--11186},
numpages = {0},
year = {1996},
publisher = {American Physical Society},
doi = {10.1103/PhysRevB.54.11169},
url = {https://link.aps.org/doi/10.1103/PhysRevB.54.11169}
}

@article{arpes_cuprates_review,
  title = {Angle-resolved photoemission studies of the cuprate superconductors},
  author = {Damascelli, Andrea and Hussain, Zahid and Shen, Zhi-Xun},
  journal = {Rev. Mod. Phys.},
  volume = {75},
  issue = {2},
  pages = {473--541},
  numpages = {0},
  year = {2003},
  month = {Apr},
  publisher = {American Physical Society},
  doi = {10.1103/RevModPhys.75.473},
  url = {https://link.aps.org/doi/10.1103/RevModPhys.75.473}
}

@article{ele_struct_cuprates_review,
  title = {Electronic structure of the high-temperature oxide superconductors},
  author = {Pickett, Warren E.},
  journal = {Rev. Mod. Phys.},
  volume = {61},
  issue = {2},
  pages = {433--512},
  numpages = {0},
  year = {1989},
  month = {Apr},
  publisher = {American Physical Society},
  doi = {10.1103/RevModPhys.61.433},
  url = {https://link.aps.org/doi/10.1103/RevModPhys.61.433}
}

@article{downfolding_hubbard_dmrg,
  title = {Density matrix renormalization group based downfolding of the three-band Hubbard model: Importance of density-assisted hopping},
  author = {Jiang, Shengtao and Scalapino, Douglas J. and White, Steven R.},
  journal = {Phys. Rev. B},
  volume = {108},
  issue = {16},
  pages = {L161111},
  numpages = {7},
  year = {2023},
  month = {Oct},
  publisher = {American Physical Society},
  doi = {10.1103/PhysRevB.108.L161111},
  url = {https://link.aps.org/doi/10.1103/PhysRevB.108.L161111}
}

@article{higashitani2013,
  author  = {Higashitani, S. and Takeuchi, H. and Matsuo, S. and Nagato, Y. and Nagai, K.},
  title   = {Magnetic Response of Odd-Frequency $s$-Wave Cooper Pairs in a Superfluid Proximity System},
  journal = {Physical Review Letters},
  volume  = {110},
  pages   = {175301},
  year    = {2013},
  doi     = {10.1103/PhysRevLett.110.175301},
  language = {english}
}

@article{higashitani2014,
  author  = {Higashitani, S.},
  title   = {Odd-frequency pairing effect on the superfluid density and the {P}auli spin susceptibility in spatially nonuniform spin-singlet superconductors},
  journal = {Physical Review B},
  volume  = {89},
  pages   = {184505},
  year    = {2014},
  doi     = {10.1103/PhysRevB.89.184505},
  language = {english}
}

@article{Dybko_2013,
doi = {10.1088/0953-8984/25/37/376001},
url = {https://doi.org/10.1088/0953-8984/25/37/376001},
year = {2013},
month = {aug},
publisher = {IOP Publishing},
volume = {25},
number = {37},
pages = {376001},
author = {Dybko, K and Aleshkevych, P and Sawicki, M and Paszkowicz, W and Przyslupski, P},
title = {The onset of ferromagnetism and superconductivity in [La0.7Sr0.3MnO3(n u.c.)/YBa2Cu3O7(2 u.c.)]20 superlattices},
journal = {Journal of Physics: Condensed Matter},
language = {english}
}

@article{selective_interlayer_coup,
  title    = "Selective interlayer ferromagnetic coupling between the Cu spins in {YBa2Cu3O7−x} grown on top of {La0.7Ca0.3MnO3}",
  author   = "Huang, S W and Wray, L Andrew and Jeng, Horng-Tay and Tra, V T and Lee, J M and Langner, M C and Chen, J M and Roy, S and Chu, Y H and Schoenlein, R W and Chuang, Y-D and Lin, J-Y",
  journal  = "Scientific Reports",
  volume   =  5,
  number   =  1,
  pages    = "16690",
  month    =  nov,
  year     =  2015,
  language = "english",
}

@article{Chaudhuri_2023,
    author = {Chaudhuri, Sayan and Chen, You-Sheng and Lin, Jauyn Grace},
    title = {Interface Effects on Magnetic Flux Pinning in La0.7Sr0.3MnO3/YBa 2Cu3O7–x Bilayers},
    journal = {ACS Omega},
    volume = {8},
    number = {19},
    pages = {16694-16699},
    year = {2023},
    month = {05},
    issn = {2470-1343},
    doi = {10.1021/acsomega.2c07928},
    url = {https://doi.org/10.1021/acsomega.2c07928},
    eprint = {https://pubs.acs.org/acsodf/article-pdf/8/19/16694/4801859/ao2c07928.pdf},
    language = {english}
}

@article{Uspenskaya_2021,
  title    = "Kinetics of Magnetization Reversal in {Superconductor--Ferromagnet} Heterostructures in Longitudinal and Perpendicular Magnetic Fields",
  author   = "Uspenskaya, L S",
  journal  = "Journal of Surface Investigation: X-ray, Synchrotron and Neutron Techniques",
  volume   =  15,
  number   =  6,
  pages    = "1159--1164",
  month    =  nov,
  year     =  2021,
  language = "english"
}

@article{Kumawat2023,
    author = {Kumawat, Sagar Mal and Dwivedi, Gopeshwar Dhar and Su, Pin Fang and Shyu, Wade Sam and Chien, Yi Hsuan and Su, Po Wei and Chung, Chia Min and Fernandez, Nestor Daniel Bermuda and Sun, Shih Jye and Hsu, Chia-Hung and Yang, Song and Chou, Hsiung},
    title = {Magnetic Field Enhancement in Critical Current and Possible Triplet Superconductivity in LSMO/YBCO/LSMO Heterostructures},
    journal = {The Journal of Physical Chemistry C},
    volume = {127},
    number = {14},
    pages = {6861-6872},
    year = {2023},
    month = {03},
    issn = {1932-7447},
    doi = {10.1021/acs.jpcc.2c08620},
    url = {https://doi.org/10.1021/acs.jpcc.2c08620},
    eprint = {https://pubs.acs.org/jpccck/article-pdf/127/14/6861/2884684/jp2c08620.pdf},
    language = {english}
}

@article{Sen_2016,
  title = {X-ray absorption study of the ferromagnetic Cu moment at the ${\mathrm{YBa}}_{2}{\mathrm{Cu}}_{3}{\mathrm{O}}_{7}/{\mathrm{La}}_{2/3}{\mathrm{Ca}}_{1/3}{\mathrm{MnO}}_{3}$ interface and variation of its exchange interaction with the Mn moment},
  author = {Sen, K. and Perret, E. and Alberca, A. and Uribe-Laverde, M. A. and Marozau, I. and Yazdi-Rizi, M. and Mallett, B. P. P. and Marsik, P. and Piamonteze, C. and Khaydukov, Y. and D\"obeli, M. and Keller, T. and Bi\v{s}kup, N. and Varela, M. and Va\v{s}\'atko, J. and Munzar, D. and Bernhard, C.},
  journal = {Phys. Rev. B},
  volume = {93},
  pages = {205131},
  year = {2016},
  doi = {10.1103/PhysRevB.93.205131},
  url = {https://doi.org/10.1103/PhysRevB.93.205131}
}

@article{Prajapat_2018,
  title = {Proximity effects across oxide-interfaces of superconductor-insulator-ferromagnet hybrid heterostructure},
  author = {Prajapat, C. L. and Singh, Surendra and Bhattacharya, D. and Ravikumar, G. and Basu, S. and Mattauch, S. and Zheng, Jian-Guo and Aoki, T. and Paul, Amitesh},
  journal = {Scientific Reports},
  volume = {8},
  pages = {3732},
  year = {2018},
  doi = {10.1038/s41598-018-22036-y},
  url = {https://doi.org/10.1038/s41598-018-22036-y}
}

@article{Hoppler_2009,
  title = {Giant superconductivity-induced modulation of the ferromagnetic magnetization in a cuprate--manganite superlattice},
  author = {Hoppler, J. and Stahn, J. and Niedermayer, Ch. and Malik, V. K. and Bouyanfif, H. and Drew, A. J. and R\"ossle, M. and Buzdin, A. and Cristiani, G. and Habermeier, H.-U. and Keimer, B. and Bernhard, C.},
  journal = {Nature Materials},
  volume = {8},
  pages = {315--319},
  year = {2009},
  doi = {10.1038/nmat2383},
  url = {https://doi.org/10.1038/nmat2383}
}

@article{Chien2013,
  title = {Visualizing short-range charge transfer at the interfaces between ferromagnetic and superconducting oxides},
  author = {Chien, Te Yu and Kourkoutis, Lena F. and Chakhalian, Jak and Gray, Benjamin and Kareev, Michael and Guisinger, Nathan P. and Muller, David A. and Freeland, John W.},
  journal = {Nature Communications},
  volume = {4},
  pages = {2336},
  year = {2013},
  doi = {10.1038/ncomms3336},
  url = {https://doi.org/10.1038/ncomms3336}
}

@article{Kim_2018_arpes,
  author  = {Kim, Chung Koo and Drozdov, Ilya K. and Fujita, Kazuhiro and Davis, J. C. S\'eamus and Bo\v{z}ovi\'c, Ivan and Valla, Tonica},
  title   = {In-situ angle-resolved photoemission spectroscopy of copper-oxide thin films synthesized by molecular beam epitaxy},
  journal = {Journal of Electron Spectroscopy and Related Phenomena},
  volume  = {257},
  pages   = {146775},
  year    = {2022},
  doi     = {10.1016/j.elspec.2018.07.003},
  url     = {https://doi.org/10.1016/j.elspec.2018.07.003}
}

@article{Zhong_2022_arpes,
  author  = {Zhong, Yong and Chen, Zhuoyu and Chen, Su-Di and Xu, Ke-Jun and Hashimoto, Makoto and He, Yu and Uchida, Shin-Ichi and Lu, Donghui and Mo, Sung-Kwan and Shen, Zhi-Xun},
  title   = {Differentiated roles of Lifshitz transition on thermodynamics and superconductivity in La$_{2-x}$Sr$_x$CuO$_4$},
  journal = {Proceedings of the National Academy of Sciences},
  volume  = {119},
  number  = {32},
  pages   = {e2204630119},
  year    = {2022},
  doi     = {10.1073/pnas.2204630119},
  url     = {https://doi.org/10.1073/pnas.2204630119}
}

@article{Iwasawa_2023_spin_arpes,
  author  = {Iwasawa, Hideaki and Sumida, Kazuki and Ishida, Shigeyuki and Le F\`evre, Patrick and Bertran, Fran\c{c}ois and Yoshida, Yoshiyuki and Eisaki, Hiroshi and Santander-Syro, Andr\'es F. and Okuda, Taichi},
  title   = {Exploring spin-polarization in Bi-based high-$T_c$ cuprates},
  journal = {Scientific Reports},
  volume  = {13},
  pages   = {13451},
  year    = {2023},
  doi     = {10.1038/s41598-023-40145-1},
  url     = {https://doi.org/10.1038/s41598-023-40145-1}
}

@article{Takigawa_1989_knight,
  author  = {Takigawa, M. and Hammel, P. C. and Heffner, R. H. and Fisk, Z.},
  title   = {Spin susceptibility in superconducting YBa$_2$Cu$_3$O$_7$ from $^{63}$Cu Knight shift},
  journal = {Physical Review B},
  volume  = {39},
  pages   = {7371--7374},
  year    = {1989},
  doi     = {10.1103/PhysRevB.39.7371},
  url     = {https://doi.org/10.1103/PhysRevB.39.7371}
}

@article{Hossain_2008_arpes,
  author  = {Hossain, M. A. and Mottershead, J. D. F. and Fournier, D. and Bostwick, A. and McChesney, J. L. and Rotenberg, E. and Liang, R. and Hardy, W. N. and Sawatzky, G. A. and Elfimov, I. S. and Bonn, D. A. and Damascelli, A.},
  title   = {In situ doping control of the surface of high-temperature superconductors},
  journal = {Nature Physics},
  volume  = {4},
  pages   = {527--531},
  year    = {2008},
  doi     = {10.1038/nphys998},
  url     = {https://doi.org/10.1038/nphys998}
}

@article{lsmo_pha_diag,
  title = {Structural, magnetic, and electrical properties of single-crystalline {L}a$_{1-x}${S}r$_{x}${M}n{O}$_{3}$ $(0.4<x<0.85)$},
  author = {Hemberger, J. and Krimmel, A. and Kurz, T. and Krug von Nidda, H.-A. and Ivanov, V. Yu. and Mukhin, A. A. and Balbashov, A. M. and Loidl, A.},
  journal = {Phys. Rev. B},
  volume = {66},
  issue = {9},
  pages = {094410},
  year = {2002},
  publisher = {American Physical Society},
  doi = {10.1103/PhysRevB.66.094410},
  url = {https://link.aps.org/doi/10.1103/PhysRevB.66.094410},
  language = {english}
}

@article{gga,
title = {Generalized {G}radient {A}pproximation {M}ade {S}imple},
author = {Perdew, John P. and Burke, Kieron and Ernzerhof, Matthias},
journal = {Phys. Rev. Lett.},
volume = {77},
issue = {18},
pages = {3865--3868},
year = {1996},
month = {Oct},
publisher = {American Physical Society},
doi = {10.1103/PhysRevLett.77.3865},
url = {https://link.aps.org/doi/10.1103/PhysRevLett.77.3865}
}

@article{paw,
  title = {From ultrasoft pseudopotentials to the projector augmented-wave method},
  author = {Kresse, G. and Joubert, D.},
  journal = {Phys. Rev. B},
  volume = {59},
  issue = {3},
  pages = {1758--1775},
  numpages = {0},
  year = {1999},
  month = {Jan},
  publisher = {American Physical Society},
  doi = {10.1103/PhysRevB.59.1758},
  url = {https://link.aps.org/doi/10.1103/PhysRevB.59.1758}
}

@Article{pseudo_test,
author={Prandini, Gianluca and Marrazzo, Antimo and Castelli, Ivano E. and Mounet, Nicolas and Marzari, Nicola},
title={Precision and efficiency in solid-state pseudopotential calculations},
journal={npj Computational Materials},
year={2018},
month={Dec},
day={06},
volume={4},
number={1},
pages={72},
issn={2057-3960},
doi={10.1038/s41524-018-0127-2},
url={https://doi.org/10.1038/s41524-018-0127-2}
}

@article{civelli_cdmft_sc_2009,
  author = {Civelli, Marcello},
  title = {Doping-driven evolution of the superconducting state from a doped {Mott} insulator: Cluster dynamical mean-field theory},
  journal = {Phys. Rev. B},
  volume = {79},
  pages = {195113},
  year = {2009},
  publisher = {American Physical Society},
  doi = {10.1103/PhysRevB.79.195113},
  url = {https://link.aps.org/doi/10.1103/PhysRevB.79.195113}
}

\end{document}